\documentclass{aa}  

\usepackage{graphicx}
\usepackage{txfonts}
\usepackage{hyperref}
\usepackage{breakurl}              
\usepackage{pdfcomment}              
\usepackage{acronym}                 
\usepackage{comment}
\usepackage{cases}
\usepackage{empheq}
\usepackage{here}
\usepackage{bm}
\usepackage[version=3]{mhchem}
\usepackage{siunitx}
\usepackage{color}

\hypersetup{colorlinks=true, linkcolor=blue, citecolor=blue}

\def\v{\upsilon}

\begin{document} 

   \title{Spin of protoplanets generated by pebble accretion: Influences of protoplanet-induced gas flow}

   \author{Kohsuke Takaoka \inst{1,2} \and Ayumu Kuwahara \inst{1,2} \and Shigeru Ida \inst{2} \and Hiroyuki Kurokawa \inst{2}}

   \institute{Department of Earth and Planetary Sciences, Tokyo Institute of Technology, Ookayama, Meguro-ku, Tokyo 152-8551, Japan \\
             \email{takaoka.k.ac@m.titech.ac.jp} \and
             Earth-Life Science Institute, Tokyo Institute of Technology, Ookayama, Meguro-ku, Tokyo 152-8550, Japan\\}

   \date{Received XXX; accepted YYY}

 
  \abstract
    {In the pebble accretion model, protoplanets accrete millimeter- to centimeter-sized particles (pebbles). When a protoplanet grows, a dense gas envelope forms around it. The envelope affects accretion of pebbles and, in particular, the spin angular momentum transfer at the collision to the planet.}
    {We investigate the spin state of a protoplanet during the pebble accretion influenced by the gas flow in the gravitational potential of the protoplanet and how it depends on the planetary mass, the headwind speed, the distance from the host star, and the pebble size.}
    {We perform nonisothermal three-dimensional hydrodynamical simulations in a local frame to obtain the gas flow around the planet. We then numerically integrate three-dimensional orbits of pebbles under the obtained gas flow. Finally, assuming uniform spatial distribution of incoming pebbles, we calculate net spin by summing up specific angular momentum that individual pebbles transfer to the protoplanet at impacts.}
    {We find that a protoplanet with the envelope acquires prograde net spin rotation regardless of the planetary mass, the pebble size, and the headwind speed of the gas. This is because accreting pebbles are dragged by the envelope that commonly has prograde rotation. As the planetary mass or orbital radius increases, the envelope is thicker and the prograde rotation is faster, resulting in faster net prograde spin. When the dimensionless thermal mass of the planet, $m = R_{\mathrm{Bondi}} / H$, where $R_{\mathrm{Bondi}}$ and $H$ are the Bondi radius and the disk gas scale height, is larger than a certain critical mass ($m \gtrsim 0.3$ at $0.1 \, \mathrm{au}$ or $m \gtrsim 0.1$ at $1 \, \mathrm{au}$), the spin rotation exceeds the breakup one.}
    {The predicted spin frequency reaches the breakup one at the planetary mass $m_{\mathrm{iso,rot}} \sim 0.1 \, (a / 1 \, \mathrm{au})^{-1/2}$ (where $a$ is the orbital radius), suggesting that the protoplanet cannot grow beyond $m_{\mathrm{iso,rot}}$. It is consistent with the Earth's current mass and could help the formation of the Moon by a giant impact on fast-spinning proto-Earth.}


   \maketitle

\section{Introduction}
\label{introduction}

    Many of solid bodies (terrestrial planets and minor bodies) in the Solar System rotate in the prograde directions, in other words, the direction of their spin coincides with the direction of their orbits \citep{Warner2009}. When targeting asteroids larger than $150 \, \mathrm{km}$ in diameter, which are considered less susceptible to post-formation dynamics and collisions \citep{Bottke2005, Steinberg2015}, and planets, their spin vectors are anisotropic and the preference of prograde spins is statistically significant \citep{Visser2020}.
    
    While gas giants generally acquire prograde spin due to gas accertion from circumplanetary disks that usually rotate in prograde directions \citep{Machida2008}, how terrestrial and icy planets acquired their spin is unclear. For instance, in the classical planetesimal accretion model, planets do not generally achieve sufficient spin angular velocities \citep{Ida1990, Lissauer1991, Lissauer1997, Dones1993a} except when the planetesimal disks have a partial gap around the protoplanet orbit \citep{Ohtsuki1998}. This is because the contribution of the planetesimals to the spin cancels out, resulting in a smaller rotation rate than observed. An alternative model for the origins of planet spins is the giant-impact model \citep{Dones1993b}, which is the current paradigm. In this approach, the rotation is dominated by a single impact of a relatively large projectile, thus the rotation is not canceled out as in the planetesimal accretion model. Indeed, planets formed predominantly through giant impacts generally have large spin frequency comparable to the breakup frequency \citep{Kokubo2007, Miguel2010}. However, this is not consistent in terms of the direction of spin axes. When a single giant impact determines the direction of rotation, the spin axis directions should follow an isotropic distribution. While the spin distribution of the terrestrial planets in the Solar System has statistical uncertainty, the preference for prograde rotation is extended down to asteroids. It is more natural to assume that there was a mechanism that made the initial rotations of solid bodies more likely to be prograde in the Solar System.
    
    In recent years, pebble accretion attracts a lot of attention as a new model of planetary formation \citep{Ormel2010, Lambrechts2012}. In this model, millimeter- to centimeter-sized particles called "pebbles," which easily couple with the gas, work as the building blocks of planets. This model, as well as the planetesimal accretion model, can explain the anisotropy of the spin vector because the averaged rotation vector has only the vertical component due to the plane symmetry of the protoplanetary disk. Thus, the question is whether pebble accretion can spin up the planet much more than planetesimal accretion.
    
    \cite{Johansen2010} were the first to investigate the spinning-up of solid bodies by pebble accretion. The authors performed hydrodynamical simulations of gas and pebbles with a $\sim 10^{2} \, \mathrm{km}$-sized solid body, in which pebbles are treated as Lagrangian particles. They considered the case that the pebbles to gas ratio is $\sim 1$. Due to the back-reaction from the pebbles through gas drag, the gas motion becomes turbulent. A prograde circumplanetary accretion disk forms, resulting in the prograde spin rotation of the accreting solid body. Although the results potentially explain the trend for the preferred prograde rotation of the Solar System bodies, they assumed a specific situation with the high pebble to gas ratio. 
    
    \cite{Visser2020} studied the link between pebble accretion and the spin rotation of solid bodies with $\sim 10 \mathrm{-} 10^{3} \, \mathrm{km}$ in diameter under a variety of disk conditions and pebble parameters. Assuming an unperturbed sub-Keplerian shear flow, the authors computed the pebble trajectories. They found that the absolute values of the net spin angular momentum of the solid bodies can be much larger in the case of pebble accretion than in the planetesimal accretion. Though in certain regions of their parameter space the net rotation can be retrograde, the contribution to the prograde rotation is dominant in most regions in their parameter space, which is consistent with observations of the spin distribution of relatively large asteroids.

    Although the results of these two previous studies are applicable when considering the spin rotation of solid bodies with $\lesssim 10^{3} \, \mathrm{km}$ in diameter, their results cannot be applied for $\gtrsim 10^{3} \, \mathrm{km}$-sized protoplanets. For larger-mass objects such as terrestrial planets in the Solar System and super-Earths, the influences of the protoplanet's gravity on the surrounding gas cannot be neglected.
    
    Recent three-dimensional (3D) hydrodynamical simulations show complex gas flows around embedded protoplanets \citep{Ormel2015b, Fung2015, Cimerman2017, Lambrechts2017, Kurokawa2018, Kuwahara2019, Bethune2019, Fung2019, Moldenhauer2021, Moldenhauer2022}. The structure of the gas flow is characterized by i) the horseshoe flow ahead and behind the protoplanet's orbital motion, ii) the streams closely passing the protoplanet with the inflow from the polar region and the outflow from the midplane region, and iii) the weak interacting streams passing in the distant region by the Keplerian shear flow. When the Bondi radius is larger than the physical radius of the protoplanet, an envelope or a primordial atmosphere is formed by the protoplanet's gravitational potential. The envelope has a higher density than in the unperturbed regions and rotates in the prograde direction due to the Coriolis force. As suggested in \cite{Kuwahara2020a, Kuwahara2020b}, the protoplanet-induced gas flow can alter the trajectories of pebbles and thus affects angular momentum transfer from the accreting pebbles.
    
    In this study, we investigate the specific angular momentum (SAM) transfer from pebbles to a protoplanet by impacts, numerically integrating the pebble trajectories one by one under the influence of the protoplanet-induced gas flow. Summing up the contributions from accreting pebbles, we derive the spin rotation state of the protoplanet as a function of the protoplanet mass, the pebble size, and the headwind speed.
    
    The structure of this paper is as follows. Our numerical models and methods are explained in Sect.~\ref{method}. We show the results of hydrodynamical simulations, orbital calculations, and spin calculations in Sect.~\ref{result}. Section~\ref{discussion} provides discussions on comparisons with observations and possible applications to planet and moon formation theory. A summary is presented in Sect.~\ref{conclusion}.

\section{Methods}
\label{method}

    In our simulation, we first perform 3D local hydrodynamical simulations of the disk gas flow under the planetary gravitational potential (Sect.~\ref{method:hydro}). Then we perform 3D orbital calculations of pebbles taking account of the aerodynamical drag from the obtained gas flow in the local coordinates co-rotating with the planet (Sect.~\ref{method:EOM}). We calculate the net SAM of the planetary spin by summing up the SAM transferred from individual pebbles at impacts (Sect.~\ref{method:momentum}).

    \subsection{Scaling}
    \label{method:scaling}
    
    \begin{figure}[tb]
        \centering
        \includegraphics[width=\linewidth, bb = 0 -1 557 470]{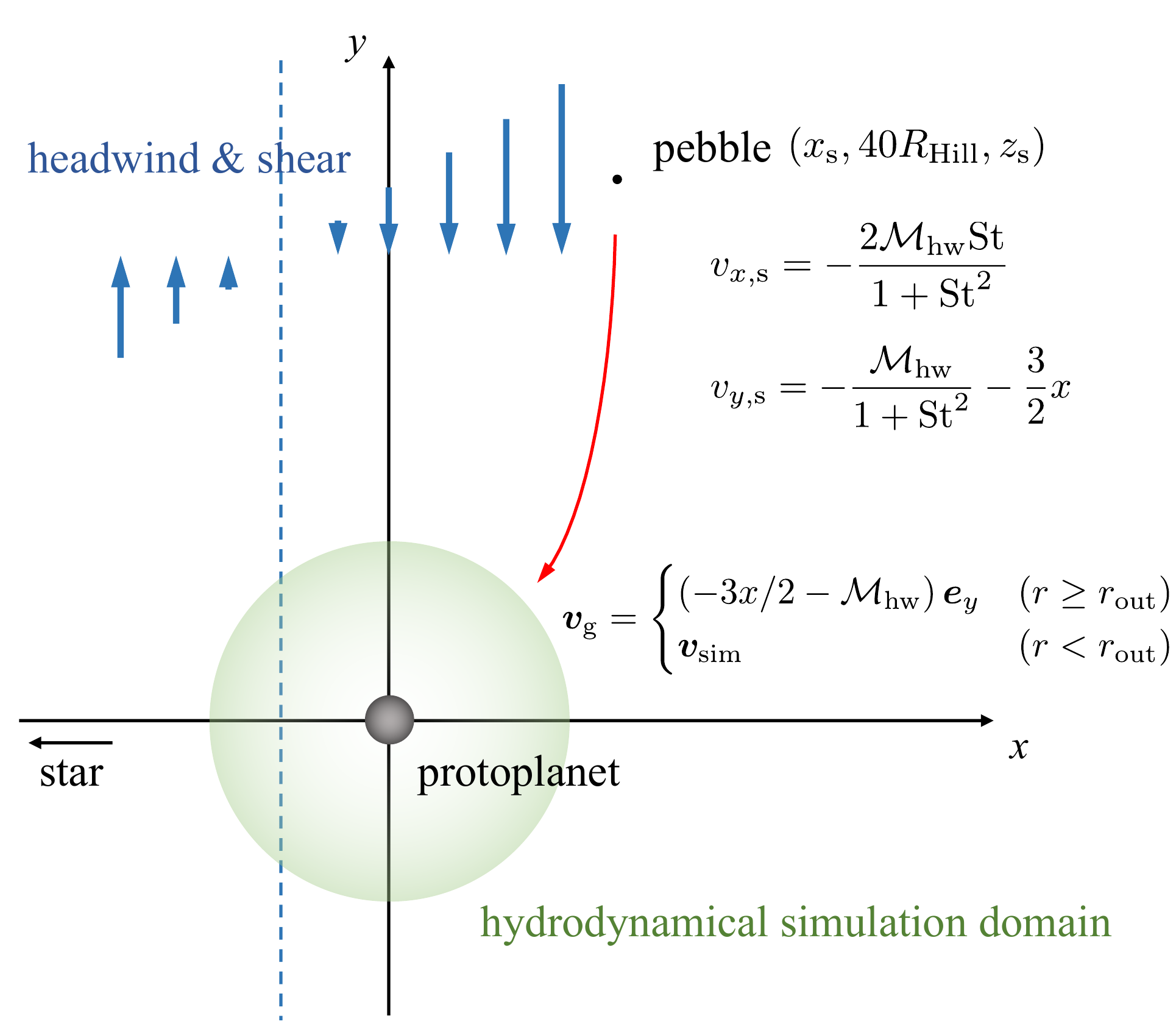}
        \caption{Schematic picture of an orbital calculation of pebbles.The starting point of pebbles is $(x_{\mathrm{s}}, |y_{\mathrm{s}}| = 40 \, R_{\mathrm{Hill}}, z_{\mathrm{s}})$. The $x$- and $z$-coordinates of the starting point of pebbles, $x_{\mathrm{s}}$ and $z_{\mathrm{s}}$, are parameters. The green circle area represents the hydrodynamical simulation domain with radius $r_{\mathrm{out}}$. We used the gas velocity obtained from the hydrodynamical simulation to calculate the gas drag force acting on the pebble within $r_{\mathrm{out}}$. Outside $r_{\mathrm{out}}$, the gas velocity is assumed to be the speed of the sub-Keplerian shear flow.}
        \label{fig:local}
    \end{figure}
    
    \begin{figure}[tb]
        \centering
        \includegraphics[width=\linewidth, bb = 0 0 720 504]{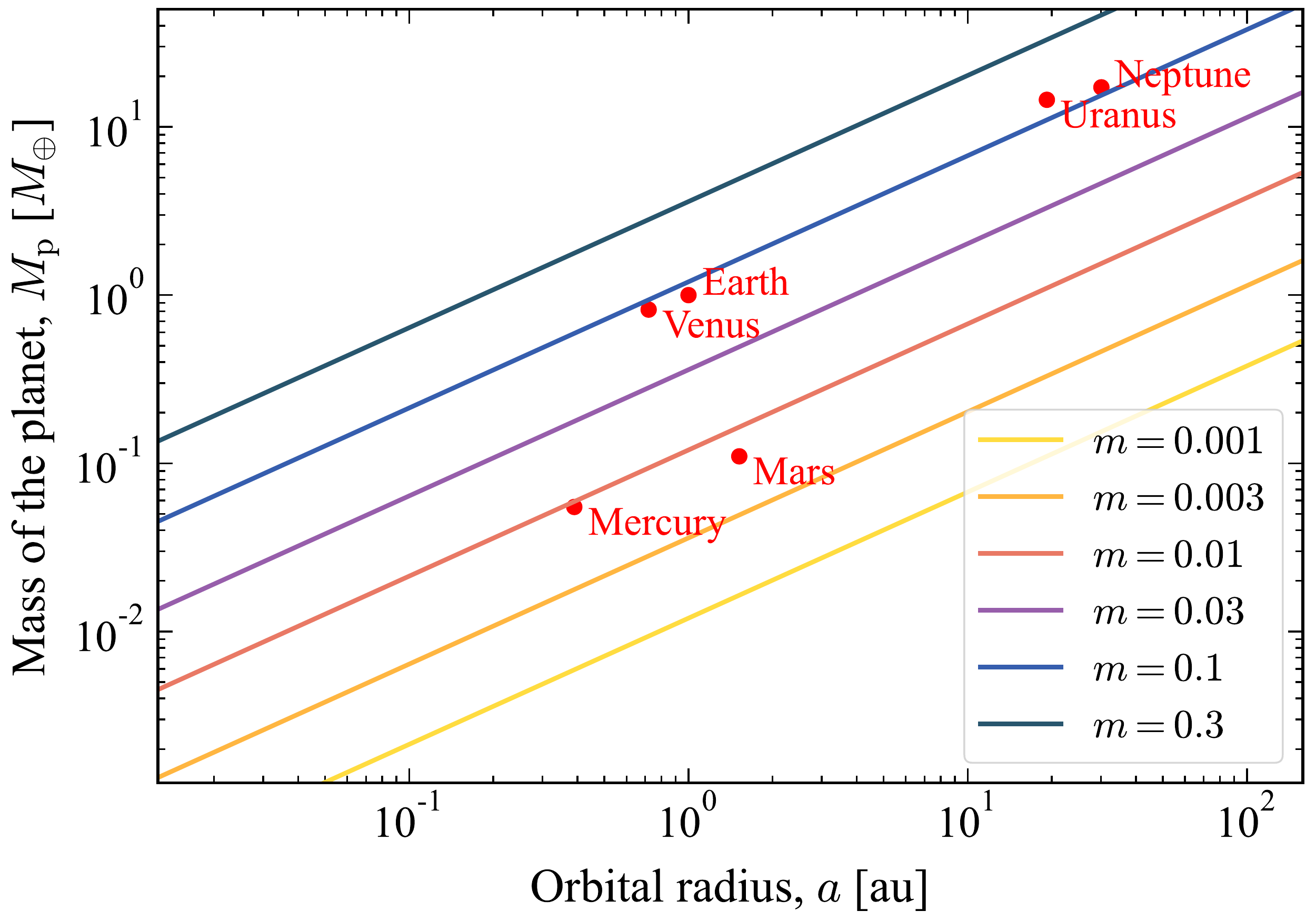}
        \caption{Relationship between dimensionless planetary masses and dimensional ones in the case of the radially optically-thin limit temperature distribution of the disks around the solar mass host star. Each line represents the dimensionless planetary mass as a function of the orbital radius. Different colors correspond to different $m$. The red dots represent the Solar System planets.}
        \label{fig:mass}
    \end{figure}
    
    In this paper, we use the local coordinates $(x, y, z)$ co-rotating with a protoplanet orbiting a host star at the radius $a$, where $x$, $y$, and $z$ are the radial, azimuthal, and vertical directions, respectively (Fig.~\ref{fig:local}). We normalize lengths of the coordinates $(x, y, z)$, times, and gas densities by the disk scale height, $H$, the reciprocal of the planetary orbital frequency, $\Omega_{\mathrm{K}}^{-1} = \sqrt{a^{3} / G M_{\ast}}$, where $G$ and $M_{\ast}$ are the gravitational constant and the host star mass, and the unperturbed gas density at $a$, $\rho_{\mathrm{disk}}$, respectively. Accordingly, velocities are normalized by the sound speed, $c_{\mathrm{s}} = H \Omega_{\mathrm{K}}$. Following \cite{Ormel2015a}, we adopt the mass scaling factor, $(H/a)^{3} M_{\ast} = \Omega_{\mathrm{K}}^{2} H^{3} / G = c_{\mathrm{s}}^{2} H / G$. In this dimensionless unit system, the dimensionless planetary mass is defined as
    \begin{equation}
    \label{eq:m}
        m \equiv \frac{M_{\mathrm{p}}}{c_{\mathrm{s}}^{2} H / G} = \frac{R_{\mathrm{Bondi}}}{H} \, ,
    \end{equation}
    where $M_{\mathrm{p}}$ is the mass of the protoplanet and $R_{\mathrm{Bondi}} = G M_{\mathrm{p}} / c_{\mathrm{s}}^{2}$ is the Bondi radius of the protoplanet, which represents a typical radius of gravitationally bound gas enveople around the protoplanet. In terms of $m$, the Hill radius is expressed by \citep{Kurokawa2018}:
    \begin{equation}
    \label{eq:Hill}
        R_{\mathrm{Hill}} = \Biggl( \frac{M_{\mathrm{p}}}{3 M_{\ast}} \Biggr)^{1/3} a = \Biggl( \frac{m}{3} \Biggr)^{1/3} H \, .
    \end{equation}
    
    To convert the normalized masses to dimensional ones, the aspect ratio of the disk, $H/a$, in other words, the disk temperature distribution needs to be specified. For simplicity, we use the radially optically thin limit, 
    \begin{align}
    \label{eq:T_thin}
        T = 270 \, \Biggl( \frac{a}{1 \, \mathrm{au}} \Biggr)^{-1/2} \, \Biggl( \frac{L_{\ast}}{1 \, L_{\odot}} \Biggr)^{1/4} \, \mathrm{K} \, ,
    \end{align}
    where $L_{\ast}$ and $L_{\odot}$ are the stellar and the Solar luminosity, respectively \citep{Hayashi1981}. With this temperature distribution, the aspect ratio is given by $H/a \simeq 0.033 \, (a / 1 \, \mathrm{au})^{1/4} \, (L_{\ast} / 1 \, L_{\odot})^{1/8}$. In this case, the dimensional planetary mass can be described by \citep{Kurokawa2018}:
    \begin{align}
    \label{eq:Mp}
        M_{\mathrm{p}} &= m \, \Biggl( \frac{H}{a} \Biggr)^{3} M_{\ast} \nonumber \\
        &\simeq 12 \, m  \,\Biggl( \frac{a}{1 \, \mathrm{au}} \Biggr)^{3/4} \Biggl( \frac{L_{\ast}}{1 \, L_{\odot}} \Biggr)^{3/8} \Biggl( \frac{M_{\ast}}{1 \, M_{\odot}} \Biggr) \, M_{\oplus} \, ,
    \end{align}
    where $M_{\odot}$ is the Solar mass. The normalized planetary masses considered in this study are $m = 0.001, 0.003, 0.01, 0.03, 0.1$, and $0.3$. Assuming $M_{\ast}= 1\, M_{\odot}$ and $L_{\ast} = 1 \, L_{\odot}$, we show the relation between $m$ and dimensional planetary masses in Fig.~\ref{fig:mass}.

    \subsection{Three-dimensional hydrodynamical simulations}
    \label{method:hydro}
    
    \begin{table*}[tb]
        \centering
        \caption{List of hydrodynamical simulations. These columns show the simulation name, the mass of the protoplanet (corresponds to the Bondi radius, see Eq.~(\ref{eq:m})), the Hill radius, the inner boundary of computational domain, the outer boundary of computational domain, the smoothing length, the Mach number of the headwind of the gas, the termination time of computation, and the dimensionless cooling time, respectively.}
        \scalebox{0.88}[0.88]{
        \begin{tabular}{lcccccccc}
            \hline
            \hline
            Name & Mass & Hill radius & Inner boundary & Outer boundary & Smoothing & Headwind & End time & Cooling time \\
            & $m \; (= R_{\mathrm{Bondi}}/H)$ & $R_{\mathrm{Hill}}/H$ & $r_{\mathrm{in}} \; (= R_{\mathrm{p}}/H)$ & $r_{\mathrm{out}}$ & $r_{\mathrm{sm}}$ & $\mathcal{M}_{\mathrm{hw}}$ & $t_{\mathrm{end}}$ & $\beta$  \\
            \hline
            \texttt{m0001-01au}      & 0.001 & 0.069 & $3    \times 10^{-3}$ & 0.05 & 0      & 0.03 & 10  & $1 \times 10^{-4}$ \\
            \texttt{m0003-01au}      & 0.003 & 0.1   & $4.33 \times 10^{-3}$ & 0.05 & 0      & 0.03 & 10  & $9 \times 10^{-4}$ \\
            \texttt{m0010-01au}      & 0.01  & 0.15  & $6.46 \times 10^{-3}$ & 0.5  & 0      & 0.03 & 50  & 0.01               \\
            \texttt{m0030-01au}      & 0.03  & 0.22  & $9.32 \times 10^{-3}$ & 0.5  & 0      & 0.03 & 50  & 0.09               \\
            \texttt{m0100-01au}      & 0.1   & 0.32  & $1.39 \times 10^{-2}$ & 5    & 0      & 0.03 & 150 & 1                  \\
            \texttt{m0300-01au}      & 0.3   & 0.46  & $2    \times 10^{-2}$ & 5    & 0      & 0.03 & 200 & 9                  \\
            \hline
            \texttt{m0100-01au-Lhw}  & 0.1   & 0.32  & $1.39 \times 10^{-2}$ & 5    & 0      & 0.01 & 150 & 1                  \\
            \texttt{m0100-01au-Hhw}  & 0.1   & 0.32  & $1.39 \times 10^{-2}$ & 5    & 0      & 0.1  & 150 & 1                  \\
            \hline
            \texttt{m0100-01au-sm01} & 0.1   & 0.32  & $1.39 \times 10^{-2}$ & 5    & $0.1m$ & 0.03 & 150 & 1                  \\
            \texttt{m0100-01au-sm02} & 0.1   & 0.32  & $1.39 \times 10^{-2}$ & 5    & $0.2m$ & 0.03 & 150 & 1                  \\
            \hline
            \texttt{m0010-01au-sm01} & 0.01  & 0.15  & $6.46 \times 10^{-3}$ & 0.5  & $0.1m$ & 0.03 & 50  & 0.01               \\
            \texttt{m0010-1au-sm01}  & 0.01  & 0.15  & $6.46 \times 10^{-4}$ & 0.5  & $0.1m$ & 0.03 & 50  & 0.01               \\
            \texttt{m0030-01au-sm01} & 0.03  & 0.22  & $9.32 \times 10^{-3}$ & 0.5  & $0.1m$ & 0.03 & 50  & 0.09               \\
            \texttt{m0030-1au-sm01}  & 0.03  & 0.22  & $9.32 \times 10^{-4}$ & 0.5  & $0.1m$ & 0.03 & 50  & 0.09               \\
            \hline
        \end{tabular}
        }
        \label{table:hydro}
    \end{table*}
    
    We perform 3D hydrodynamical simulations of the gas influenced by the gravity of the protoplanet until the system reaches a steady state, and used these results in the following pebble trajectory calculations (Sect.~\ref{method:EOM}). Our methods of hydrodynamical simulations are the same as \cite{Kurokawa2018} and \cite{Kuwahara2020a, Kuwahara2020b}, except for the ranges of the planetary masses, the smoothing lengths, and the sizes of the inner boundary. We perform hydrodynamical simulations in spherical polar coordinates centered at the protoplanet, assuming a compressible and inviscid fluid of an ideal gas. We use the Athena++ code\footnote{https://github.com/PrincetonUniversity/athena} \citep{White2016, Stone2020}. The parameters of hydrodynamical simulations are summarized in Table~\ref{table:hydro}.
    
    We set an inner boundary of the computational domain at the normalized physical radius of the protoplanet, $r_{\mathrm{in}} = R_{\mathrm{p}}/H$, where $R_{\mathrm{p}}$ is the dimensional physical radius:
    \begin{align}
    \label{eq:Rp1}
        R_{\mathrm{p}} & \simeq 3 \times 10^{-3} \, m^{1/3} \, \Biggl( \frac{\rho_{\mathrm{p}}}{5 \, \mathrm{g} \, \mathrm{cm}^{-3}} \Biggr)^{-1/3} \Biggl( \frac{M_{\ast}}{1 \, M_{\odot}} \Biggr)^{1/3} \, \Biggl( \frac{a}{1 \, \mathrm{au}} \Biggr)^{-1} \, H \\
    \label{eq:Rp2}
        & \simeq 4.4 \times 10^{-3} \, \Biggl( \frac{\rho_{\mathrm{p}}}{5 \, \mathrm{g} \, \mathrm{cm}^{-3}} \Biggr)^{-1/3} \Biggl( \frac{M_{\ast}}{1 \, M_{\odot}} \Biggr)^{1/3} \, \Biggl( \frac{a}{1 \, \mathrm{au}} \Biggr)^{-1} \, R_{\mathrm{Hill}} \, ,
    \end{align}
    where $\rho_{\mathrm{p}}$ is the bulk density of the protoplanet \citep{Kuwahara2020a, Kuwahara2020b}. We perform the simulations with  $r_{\mathrm{in}} = 3 \times 10^{-2} \, m^{1/3}$ and $3 \times 10^{-3} \, m^{1/3}$, corresponding to $a = 0.1 \, \mathrm{au}$ and $1 \, \mathrm{au}$ in the case of $M_{\ast} = 1\, M_{\odot}$ and $\rho_{\mathrm{p}} = 5 \, \mathrm{g} \, \mathrm{cm}^{-3}$ (Eq.~(\ref{eq:Rp1}); for details, see the results in Sect.~\ref{result:dependence_au}). 

    The nondimensional gas velocity at the outer boundary of the hydrodynamical simulation domain, $r_{\mathrm{out}}$, is set by the unperturbed gas velocity in the local frame,
    \begin{equation}
    \label{eq:shear}
        \bm{\v}_{\mathrm{g}, \infty}(x) = \Biggl( - \frac{3}{2} x - \mathcal{M}_{\mathrm{hw}} \Biggr) \, \bm{e}_{y} \, ,
    \end{equation}
    where $\mathcal{M}_{\mathrm{hw}} = \v_{\mathrm{hw}} / c_{\mathrm{s}}$ is the Mach number of the headwind of the gas. The headwind velocity, $\v_{\mathrm{hw}}$, represents the deviation from the Keplerian velocity due to the radial pressure gradient, 
    \begin{equation}
    \label{eq:headwind}
        \v_{\mathrm{hw}} = \eta \, \v_{\mathrm{K}} \, ; \ \ \eta = - \frac{1}{2} \, \Biggl( \frac{c_{\mathrm{s}}}{\v_{\mathrm{K}}} \Biggr)^{2} \, \frac{\partial \ln P}{\partial \ln a} \, ,
    \end{equation}
    where $\v_{\mathrm{K}} = a \Omega_{\mathrm{K}}$ is the Keplerian velocity and $P$ is the pressure of the gas. In the optically-thin limit disks, the Mach number of the headwind is $\mathcal{M}_{\mathrm{hw}} \simeq 0.05 \, (a / 1 \, \mathrm{au})^{1/4}$. We consider $\mathcal{M}_{\mathrm{hw}} = 0.03$ as a fiducial parameter (Table~\ref{table:hydro}). The dependence on the Mach number of the headwind is investigated in Sect.~\ref{result:dependence_hw}.
    
    The dimensionless continuity, Euler's, and energy conservation equations are described as follows:
    \begin{gather}
    \label{eq:mass_const}
        \frac{\partial \rho_{\mathrm{g}}}{\partial t} + \nabla \cdot (\rho_{\mathrm{g}} \bm{\v}_{\mathrm{g}}) = 0 \, , \\[8pt]
    \label{eq:momentum_const}
        \Biggl( \frac{\partial}{\partial t} + \bm{\v}_{\mathrm{g}} \cdot \nabla \Biggr) \, \bm{\v}_{\mathrm{g}} = - \frac{\nabla P}{\rho_{\mathrm{g}}} + \bm{F}_{\mathrm{cor}} + \bm{F}_{\mathrm{tid}} + \bm{F}_{\mathrm{hw}} + \bm{F}_{\mathrm{p}} \, , \\[8pt]
    \label{eq:energy_const}
        \begin{split}
            \frac{\partial E}{\partial t} + \nabla \cdot \left[ \bm{\v}_{\mathrm{g}} (E + P) \right] = \; & \rho_{\mathrm{g}} \bm{\v}_{\mathrm{g}} \cdot (\bm{F}_{\mathrm{cor}} + \bm{F}_{\mathrm{tid}} + \bm{F}_{\mathrm{hw}} + \bm{F}_{\mathrm{p}}) \\
            & - \frac{U(\rho_{\mathrm{g}} ,T) - U(\rho_{\mathrm{g}} , T_{0})}{\beta} \, ,
        \end{split}
    \end{gather}
    where $\rho_{\mathrm{g}}$ is the gas density. The internal energy density, $U$, and the total energy density, $E$, are given by
    \begin{gather}
        U = \frac{P}{\gamma - 1} \, , \\[2pt]
        E = U + \frac{1}{2} \rho_{\mathrm{g}} \v_{\mathrm{g}}^{2} \, ,
    \end{gather}
    where $\gamma$ is the specific heat ratio. We assume $\gamma = 1.4$. The last term in Eq.~(\ref{eq:energy_const}) is the radiative cooling implemented by using the $\beta$ cooling model, in which the temperature $T$ varies on the dimensionless timescale $\beta$ toward the background temperature $T_{0}$ \cite[e.g.,][]{Gammie2001}. We adopt $\beta = (m / 0.1)^{2}$ \citep{Kurokawa2018}.
    
    The right-hand sides of Eq.~(\ref{eq:momentum_const}) and (\ref{eq:energy_const}) include external force terms; the Coriolis force, $\bm{F}_{\mathrm{cor}} = -2 \bm{e}_{z} \times \bm{\v}_{\mathrm{g}}$, the tidal force, $\bm{F}_{\mathrm{tid}} = 3x \bm{e}_{x} - z \bm{e}_{z}$, and the global pressure force due to the sub-Keplerian motion of the gas, $\bm{F}_{\mathrm{hw}} = -2 \mathcal{M}_{\mathrm{hw}} \bm{e}_{x}$. The protoplanet gravitional force $\bm{F}_{\mathrm{p}}$ is given by \citep{Ormel2015b}:
    \begin{equation}
    \label{eq:smoothing}
        \bm{F}_{\mathrm{p}} = - \nabla \, \left( \frac{m}{\sqrt{r^{2} + r_{\mathrm{sm}}^{2}}} \right) \left\{ 1 - \exp \left[ - \frac{1}{2} \left( \frac{t}{t_\mathrm{inj}} \right)^{2} \right] \right\} \, ,
    \end{equation}
    where $r = \sqrt{x^{2} + y^{2} + z^{2}}$ is the distance from the protoplanet and $r_{\mathrm{sm}}$ is the normalized smoothing length. We assume $r_{\mathrm{sm}} = 0$ for most of the simulations, but we also test the cases of $r_{\mathrm{sm}} = 0.1 \, m$ and $r_{\mathrm{sm}} = 0.2 \, m$ in several runs (Table~\ref{table:hydro} and Sect.~\ref{result:dependence_rsm}). Following \citet{Ormel2015a}, we gradually inserted the protoplanet gravity at the injection time, $t_{\mathrm{inj}} = 0.5$, in order to avoid shock formation. The resolution of our simulations are 128 logarithmically spaced cells in the radial, 64 cells in the polar, and 128 cells in the azimuthal direction.

    \subsection{Three-dimensional orbital calculation of pebbles}
    \label{method:EOM}
    
    Following \cite{Kuwahara2020a, Kuwahara2020b}, we calculate the trajectories of pebbles influenced by the protoplanet-induced gas flow in the frame co-rotating with the protoplanet (Fig.~\ref{fig:local}), using a fifth-order Runge-Kutta-Fehlberg variable step scheme \citep[RKF45;][]{Fehlberg1969}. The dimensionless equation of motion for a pebble with position $\bm{r} = (x, y, z)$ and velocity $\bm{\v} = (\v_{x}, \v_{y}, \v_{z})$ are described by \citep{Ormel2010, Visser2020, Kuwahara2020a, Kuwahara2020b}:
    \begin{equation}
    \label{eq:EOM}
        \frac{\mathrm{d} \bm{\v}}{\mathrm{d} t} =
        \begin{pmatrix}
        2 \v_{y} + 3 x \\
        -2 \v_{x} \\
        0
        \end{pmatrix}
        - \frac{m}{r^{3}}
        \begin{pmatrix}
            x \\
            y \\
            z
        \end{pmatrix}
        - \frac{\bm{\v} - \bm{\v}_{\mathrm{g}}}{\mathrm{St}} \, .
    \end{equation}
    The first to third terms on the right-hand side of Eq.~(\ref{eq:EOM}) are the Coriolis and tidal forces, the gravitaional force of the protoplanet, and the gas drag force acting on the pebble, respectively. Assuming the balance between the turbulent diffusion and the vertical tidal force, we omit the $z$-component of the tidal force, $-z \bm{e}_{z}$, in Eq.~(\ref{eq:EOM}) \citep{Kuwahara2020a}. The dimensionless stopping time, called the Stokes number, is defined by
    \begin{equation}
    \label{eq:St}
        \mathrm{St} = t_{\mathrm{stop}} \, \Omega_{\mathrm{K}} \, , 
    \end{equation} 
    where the stopping time, $t_{\mathrm{stop}}$, for a pebble with the physical radius, $s$, and the internal density, $\rho_{\bullet}$, is given by
    \begin{equation}
    \label{eq:t_stop}
        t_{\mathrm{stop}} =
        \begin{cases}
            \displaystyle \frac{\rho_{\bullet} s}{\rho_{\mathrm{g}} c_{\mathrm{s}}} & \displaystyle \Biggl( \text{Epstein regime:} \ s < \frac{9}{4} l_{\mathrm{mfp}} \Biggr) \\[10pt]
            \displaystyle \frac{4 \rho_{\bullet} s^{2}}{9 \rho_{\mathrm{g}} c_{\mathrm{s}} l_{\mathrm{mfp}}} & \displaystyle \Biggl( \text{Stokes regime:} \ s \geq \frac{9}{4} l_{\mathrm{mfp}} \Biggr)
        \end{cases}
        \, .
    \end{equation}
    The mean free path of the gas is given by $l_{\mathrm{mfp}} = \mu m_{\mathrm{H}} / \rho_{\mathrm{g}} \sigma_{\mathrm{mol}} = 1.44 \, (a / 1 \, \mathrm{au})^{11/4}\, \mathrm{cm}$ for the Minimum-Mass Solar Nebula (MMSN) model \citep{Hayashi1981} with $\mu$, $m_{\mathrm{H}}$, and $\sigma_{\mathrm{mol}}$ being the mean molecular weight, $\mu = 2.34$, the mass of a proton, and the molecular collision cross section, $\sigma_{\mathrm{mol}} = 2 \times 10^{-15} \, \mathrm{cm}^{2}$ \citep{Chapman1970, Weidenschilling1977a, Nakagawa1986}. We assume $\mathrm{St} = 10^{-3} \mathrm{-} 10^{0}$ as the initial Stokes number.
    
    As shown in Eq.~(\ref{eq:t_stop}), the gas drag law changes at $s \sim l_{\mathrm{mfp}}$. Since $l_{\mathrm{mfp}} \propto \rho_{\mathrm{g}}^{-1}$, when the disk gas density is low enough to satisfy $s \lesssim l_{\mathrm{mfp}}$, the drag force increases with $\rho_{\mathrm{g}}$. In other words, $\mathrm{St}$ is inversely proportional to $\rho_{\mathrm{g}}$. This drag regime is called the Epstein regime. As a pebble approaches a protoplanet and enters its gas envelope, the drag force is stronger ($\mathrm{St}$ becomes smaller) and the pebble is more susceptible to gas drag. When $\rho_{\mathrm{g}}$ becomes high enough to satisfy $s \gtrsim l_{\mathrm{mfp}}$, the drag law is switched to the Stokes regime. Since the Stokes drag force is independent of $\rho_{\mathrm{g}}$ ($l_{\mathrm{mfp}} \propto \rho_{\mathrm{g}}^{-1}$), the Stokes number remains constant once the gas drag law switches to the Stokes regime.
    
    If $\rho_{\mathrm{g}}$ already satisfies the condition for the Stokes regime in the regions far from the protoplanet, for example in the inner region of the disk \citep[$\lesssim 1 \, \mathrm{au}$ for the MMSN model;][]{Lambrechts2012}, $\mathrm{St}$ is kept constant even when the pebble enters the gas envelope of the protoplanet. In this case, we do not need to specify the background gas density. On the other hand, in the case starting from the Epstein regime, $\mathrm{St}$ varies along the pebble trajectory and thus the background gas density needs to be specified to identify where the gas drag regime changes.
    
    We investigate two limiting cases, where either the Stokes or the Epstein regime is only adopted, and bracket intermediate cases for simplicity. This approach leaves the background gas density as a free parameter and thus we do not lose generality. As shown later in Sect.~\ref{result:net_SAM}, the net SAM transferred to the protoplanet is similar between these two limiting cases as a function of $\mathrm{St}$, implying that the rate of the angular momentum transfer obtained in this study is robust.

    The gas velocity, $\bm{\upsilon}_{\mathrm{g}}$, is switched at the outer boundary of the hydrodynamical simulation domain, $r_{\mathrm{out}}$, \citep{Kuwahara2020a, Kuwahara2020b}:
    \begin{equation} 
        \displaystyle \bm{\v}_{\mathrm{g}} = 
        \begin{cases}
            \displaystyle \left( - \frac{3}{2} x - \mathcal{M}_{\mathrm{hw}} \right) \, \bm{e}_{y} & \left( r \geq r_{\mathrm{out}} \right) \\
            \displaystyle \bm{\v}_{\mathrm{sim}} & \left( r < r_{\mathrm{out}} \right)
        \end{cases} \, ,
    \end{equation}
    where $\bm{\upsilon}_{\mathrm{sim}}$ is the gas velocity obtained from the hydrodynamical simulations.\footnote{For $m = 0.03$ and $0.1$, to avoid numerically artificial vortices, we use a smaller simulation domain of $r < 0.6 \, r_{\mathrm{out}}$ (see \citealp{Kuwahara2020a, Kuwahara2020b}, for the discussion).} We confirmed that the connection at $r = r_{\mathrm{out}}$ has only minor effect on trajectories of pebbles; the velocities of pebbles at the boundary are nearly identical to their unperturbed velocities \citep[see Eqs.~(\ref{eq:vx_s}) and (\ref{eq:vy_s}) given below;][]{Kuwahara2020a, Kuwahara2020b, Kuwahara2022}.
    
    The normalized equation of motion (Eq.~(\ref{eq:EOM})) is formally characterized by only two parameters, $m$ and $\mathrm{St}$ defined by Eqs.~(\ref{eq:m}) and (\ref{eq:St}). However, the gas flow field ($\bm{\v}_{\mathrm{g}}$) depends on $R_{\mathrm{p}}/H$ and accordingly on $a$ (Eq.~(\ref{eq:Rp1})), in particular within the Bondi radius. We will discuss the $a$-dependence in Sect.~\ref{result:dependence_au}.

    Initial conditions of the orbital integrations are given as follows. We adopt the initial $y$ as $|y_{\mathrm{s}}| = 40 \, R_{\mathrm{Hill}}/H$ \citep{Ida1989}. We integrate the orbits of pebbles in the ranges of the initial $x$ and $z$ ($x_{\mathrm{s}}$ and $z_{\mathrm{s}}$) broad enough to cover all the collision bands with the protoplanet. We resolve the collision bands with the interval of $x_{\mathrm{s}}$ as $\Delta x_{\mathrm{s}} = 0.002 \, w_{\mathrm{acc}}(0)$, which have on the order of $10^{-5} \mathrm{-} 10^{-3} \, H$. The width of the accretion window is defined as
    \begin{equation}
        w_{\mathrm{acc}}(z) \equiv x_{\mathrm{max}}(z) - x_{\mathrm{min}}(z) \, ,
    \end{equation}
    where $x_{\mathrm{max}}$ and $x_{\mathrm{min}}$ are the two ends of the collision band at a given $z$ \citep{Kuwahara2020a, Kuwahara2020b}. Although the SAM transfer to the protoplanet sensitively depends of $x_{\mathrm{s}}$ (described later in Sect.~\ref{result:individual_SAM}), we ensure that this interval has high enough resolution to estimate the net SAM. We also resolve the collision bands in the vertical direction with $\Delta z_{\mathrm{s}} = 0.05 \, b_{x}$, where $b_{x}$ is the maximum impact parameter of accreted pebbles in the unperturbed flow \citep{Ormel2012}:
    \begin{equation}
    \label{eq:bx}
        b_{x} = b_{x,0} \exp \left[ - \, \Biggl( \frac{\mathrm{St}}{2} \Biggr)^{0.65} \right] \, ,
    \end{equation}
    where $b_{x, 0}$ is described as \citep{Ormel2010, Lambrechts2012, Guillot2014, Ida2016, Sato2016}:
    \begin{equation}
    \label{eq:b0}
        b_{x,0} \simeq \min \, \left( 2 \sqrt{\frac{m \, \mathrm{St}}{\mathcal{M}_{\mathrm{hw}}}}, \, 2 \, \left( \frac{m \, \mathrm{St}}{3} \right)^{1/3} \right) \, .
    \end{equation}
    Owing to the symmetry of the system, we only consider $z_{\mathrm{s}} \geq 0$.
    
    The $x$- and $y$-components of the initial velocity of the pebble are given by the drift equations \citep{Weidenschilling1977b, Nakagawa1986}:
    \begin{gather}
    \label{eq:vx_s}
        \v_{x, \mathrm{s}} = - \frac{2 \mathcal{M}_{\mathrm{hw}} \mathrm{St}}{1 + \mathrm{St}^{2}} \, , \\[6pt]
    \label{eq:vy_s}
        \v_{y, \mathrm{s}} = - \frac{\mathcal{M}_{\mathrm{hw}}}{1 + \mathrm{St}^{2}} - \frac{3}{2} x \, .
    \end{gather}
    The initial velocity of the pebble in the vertical direction is $0$.
    
    The orbital calculation ends when a pebble reaches the protoplanet's surface ($r < r_{\mathrm{in}}$) or leaves the computational domain ($|y| > 40 \, R_{\mathrm{Hill}}/H$). However, we find that these termination conditions are not sufficient because some pebbles are trapped in the horseshoe region or continue to circulate the protoplanet. To reduce the computational time, we add the additional termination conditions for these cases. In the former case, the calculation is terminated after the $y$-directional turns outside the Hill radius are detected five times. In the latter case, the calculation is terminated after the $y$-directional turns inside the Bondi radius are detected 50 times and the effective Stokes number falls below $\mathrm{St} \, (\rho_{\mathrm{g}} / \rho_{\mathrm{g,\infty}})^{-1} \leq 10^{-4}$ in the Epstein drag case, where $\rho_{\mathrm{g,\infty}}$ is the gas density outside the computational domain.

    \subsection{Pebble-to-planet angular momentum transfer}
    \label{method:momentum}
    
    When a pebble collides with the protoplanet, we assume that the impact angular momentum is 100\% transferred to the protoplanet's spin. Because of the plane symmetry of the disk, the $x$- and $y$-components of the cumulative angular momentum should cancel each other out. Thus we compute only the $z$-component of the impact SAM, which is given by \citep{Dones1993a}:
    \begin{equation}
    \label{eq:SAM}
        l_{z} = (x \v_{y} - y \v_{x}) + \Omega_{\mathrm{K}}(x^{2} + y^{2}) \, ,
    \end{equation}
    where the second term in the right-hand side of Eq.~(\ref{eq:SAM}) is a correction term due to the co-rotating frame \citep[e.g.,][]{Dones1993a}. In general, the first term can be either positive or negative, while the second term is always positive. The first and second terms in Eq.~(\ref{eq:SAM}) for a single impact are $\sim \sqrt{2GM_{\mathrm{p}} R_{\mathrm{p}}}$ and $\sim \sqrt{GM_{\ast} / a^{3}} \times R_{\mathrm{p}}^{2}$, respectively. When the $z$-component of the angular momentum does not significantly cancel over many pebble impacts, the ratio of the second term to the first term in Eq.~(\ref{eq:SAM}) is $ \sim \sqrt{(M_{\ast} / M_{\mathrm{p}}) \, (R_{\mathrm{p}} / a)^{3}} \sim (R_{\mathrm{p}} / R_{\mathrm{Hill}})^{3/2} \ll 1$ (Eq.~(\ref{eq:Rp2})). Then the second term contribution is negligible. The net SAM transferred to the protoplanet, $\langle l_{z} \rangle$, is calculated by \citep{Dones1993a}:
    \begin{equation} 
    \label{eq:l_z}
        \langle l_{z} \rangle = \frac{\displaystyle \iint_{\mathrm{acc}} F(x_{\mathrm{s}}, z_{\mathrm{s}}) l_{z}(x_{\mathrm{s}}, z_{\mathrm{s}}) \mathrm{d} x \mathrm{d} z}{\displaystyle \iint_{\mathrm{acc}} F(x_{\mathrm{s}}, z_{\mathrm{s}}) \mathrm{d} x \mathrm{d} z} \, ,
    \end{equation}
    where the flux of pebbles entering the collision bands, $F(x_{\mathrm{s}}, z_{\mathrm{s}})$, is given by
    \begin{align}
    \label{eq:flux}
        F(x_{\mathrm{s}}, z_{\mathrm{s}}) = \left\lvert \, \v_{y, \mathrm{s}}(x_{\mathrm{s}}) \, \right\rvert \, \rho_{\mathrm{peb}}(z_{\mathrm{s}}) = \left\lvert \, - \frac{\mathcal{M}_{\mathrm{hw}}}{1 + \mathrm{St}^{2}} - \frac{3}{2} x_{\mathrm{s}} \, \right\rvert \, \rho_{\mathrm{peb}}(z_{\mathrm{s}}) \, .
    \end{align}
    The pebble density distribution in the $z$-direction, $\rho_{\mathrm{peb}}$, is assumed to be
    \begin{equation}
    \label{eq:rho_peb}
        \rho_{\mathrm{peb}} (z) = \frac{\Sigma_{\mathrm{peb}}}{\sqrt{2 \pi} H_{\mathrm{peb}}} \exp{\left[ - \frac{1}{2} \, \Biggl( \frac{z}{H_{\mathrm{peb}}} \Biggr)^{2} \right]} \, ,
    \end{equation}
    where $\Sigma_{\mathrm{peb}}$ is the surface density of pebbles in the disk and $H_{\mathrm{peb}}$ is the scale height of pebbles \citep{Dubrulle1995, Cuzzi1993, Youdin2007}:
    \begin{equation}
    \label{eq:H_peb}
        H_{\mathrm{peb}} = \Biggl( 1 + \frac{\mathrm{St}}{\alpha} \frac{1 + 2 \mathrm{St}}{1 + \mathrm{St}} \Biggr)^{-1/2} H \, ,
    \end{equation}
    where $\alpha$ is the dimensionless turbulent parameter in the disk \citep{Shakura1973}. We assume $\alpha = 10^{-3}$. Because $\Sigma_{\mathrm{peb}}$ terms in the numerator and denominator in Eq.~(\ref{eq:l_z}) cancel each other out, we do not need to give a specific value of $\Sigma_{\mathrm{peb}}$.
    
    When the mass and the physical radius of the planet are $M_{\mathrm{p}}'$ and $R_{\mathrm{p}}'$, the SAM for a grazing impact with free-fall velocity is $l_{z, \mathrm{esc}}(R_{\mathrm{p}}') = \sqrt{2GM_{\mathrm{p}}' R_{\mathrm{p}}'}$. Assuming that the planet bulk density $\rho_{\mathrm{p}}$ is constant, $l_{z, \mathrm{esc}}$ is proportional to $l_{z, \mathrm{esc}} \propto \sqrt{M_{\mathrm{p}}' R_{\mathrm{p}}'} \propto R_{\mathrm{p}}'^{2}$. When the net specific mean angular momentum, $\langle l_{z}(R_{\mathrm{p}}') \rangle$, is always proportional to $l_{z, \mathrm{esc}}(R_{\mathrm{p}}')$, namely when $C = \langle l_{z}(R_{\mathrm{p}}') \rangle \, / \, l_{z, \mathrm{esc}}(R_{\mathrm{p}}')$ is a constant, the spin angular momentum acquired during the growth of the protoplanet up to the physical radius $R_{\mathrm{p}}$ is given by
    \begin{align}
    \label{eq:L1}
        L & = \int_{0}^{R_{\mathrm{p}}} \langle l_{z}(R_{\mathrm{p}}^{\prime}) \rangle \times 4 \pi R_{\mathrm{p}}'^{2} \, \rho_{\mathrm{p}} \, \mathrm{d} R_{\mathrm{p}}' \nonumber \\
        &= 4 \pi \rho_{\mathrm{p}} C \int_{0}^{R_{\mathrm{p}}} l_{z, \mathrm{esc}}(R_{\mathrm{p}}') \, R_{\mathrm{p}}'^{2} \, \mathrm{d} R_{\mathrm{p}}' \nonumber \\[2pt]
        &= 4 \pi \rho_{\mathrm{p}} C \times \frac{1}{5} \, l_{z, \mathrm{esc}}(R_{\mathrm{p}}) \, R_{\mathrm{p}}^{3} \nonumber \\[4pt]
        & = \frac{3}{5} \, \frac{\langle l_{z}(R_{\mathrm{p}}) \rangle}{l_{z, \mathrm{esc}}(R_{\mathrm{p}})} \, M_{\mathrm{p}} \, l_{z, \mathrm{esc}}(R_{\mathrm{p}}) \nonumber \\[4pt]
        &= \frac{3\sqrt{2}}{5} \, \frac{\langle l_{z}(R_{\mathrm{p}}) \rangle}{l_{z, \mathrm{esc}}(R_{\mathrm{p}})}\, M_{\mathrm{p}} \, R_{\mathrm{p}}^{2} \, \omega_{\mathrm{crit}} \, ,
    \end{align}
    where $\omega_{\mathrm{crit}} \equiv \sqrt{G M_{\mathrm{p}} / R_{\mathrm{p}}^{3}}$ is the breakup frequency. Using the spin angular velocity $\omega$ at $R_{\mathrm{p}}$ and assuming that the protoplanet is a sphere of a uniform density, the spin angular momentum is expressed by
    \begin{align}
    \label{eq:L2}
        L = \frac{2}{5} \, M_{\mathrm{p}} R_{\mathrm{p}}^{2} \,\omega \, .
    \end{align}
    From Eqs.~(\ref{eq:L1}) and (\ref{eq:L2}), we obtain
    \begin{equation}
    \label{eq:breakup}
       \frac{\omega}{\omega_{\mathrm{crit}}} = \frac{3}{\sqrt{2}} \frac{\langle l_{z} \rangle}{l_{z, \mathrm{esc}}} \, ,
    \end{equation}
    at $R_{\mathrm{p}}$. This means that the protoplanet's rotation reaches the breakup frequency if pebbles constantly transfer $\langle l_{z} \rangle \gtrsim 0.47 \, l_{z, \mathrm{esc}}$ on average to the protoplanet in the course of its growth.

\section{Results}
\label{result}

    In this section, we present the results obtained by the numerical methods described in Sect.~\ref{method}. First, we show the results of hydrodynamical simulations in Sect.~\ref{result:hydro}. In Sect.~\ref{result:trajectory}, we show the trajectories of individual pebbles. Section~\ref{result:individual_SAM} shows the SAM transferred by the individual pebbles to the protoplanet. We present how much net SAM the protoplanet acquires as a function of the dimensionless planetary mass and Stokes number of the pebbles in Sect.~\ref{result:net_SAM}, which is our main results. Sections~\ref{result:dependence_hw}--\ref{result:dependence_au} show the dependence on the headwind speed, the smoothing length, and the orbital radius, respectively.

    \subsection{Protoplanet-induced gas flow}
    \label{result:hydro}
    
    \begin{figure*}[tb]
        \centering
        \includegraphics[width=0.9\linewidth, bb = 13 11 994 413]{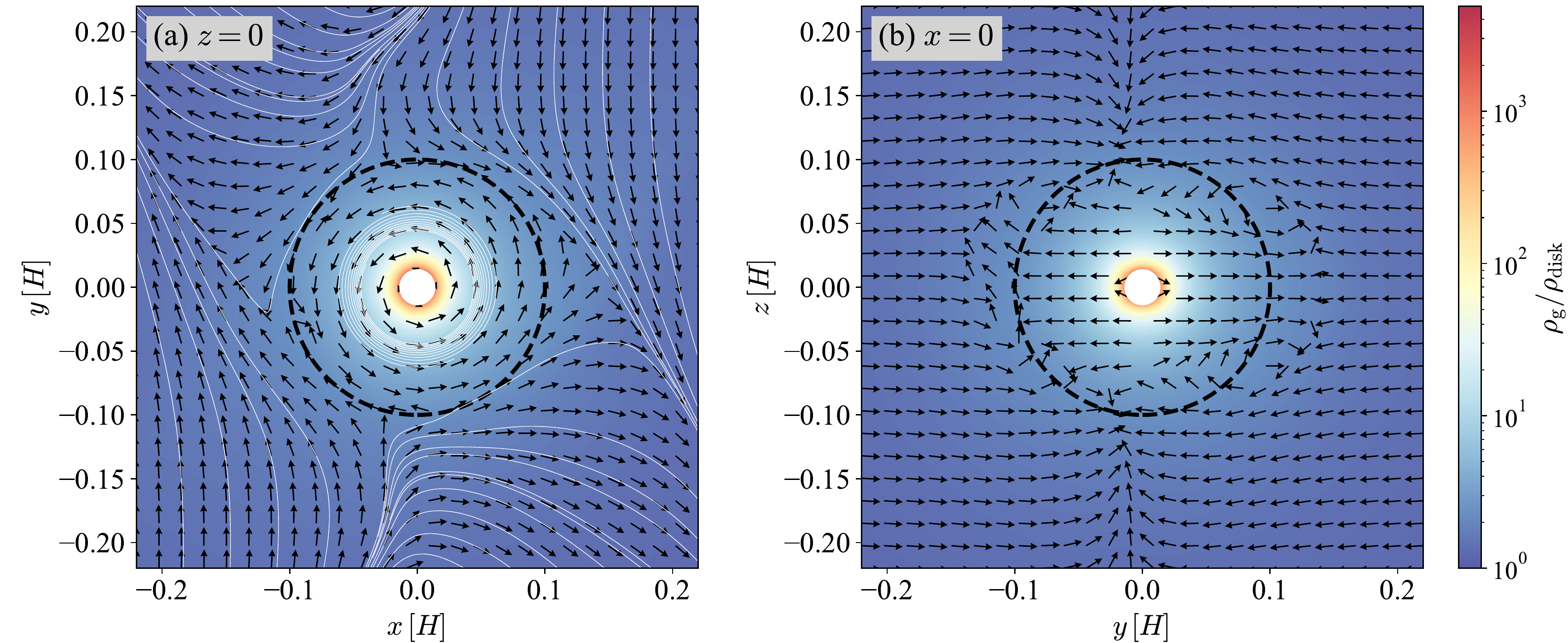}
        \caption{Flow structure around a protoplanet at (a) $z =0$ (the midplane of the disk) and (b) $x = 0$. The result is obtained from the \texttt{m0100-01au} run. The white-filled circle at the center represents the protoplanet and the black dashed circle represents the Bondi radius. The black arrows represent the direction of the gas flow. The white solid lines represent the gas streamlines. The color contours represent the gas density.}
        \label{fig:hydro}
    \end{figure*}
    
    Figure~\ref{fig:hydro}a shows an example of a protoplanet-induced flow at the midplane. The result is obtained from the \texttt{m0100-01au} run. The Mach number of the headwind of the gas is $\mathcal{M}_{\mathrm{hw}} = 0.03$. As shown in previous studies \citep{Ormel2015b, Kuwahara2019}, protoplanet-induced gas flow near the midplane are characterized by the shear streamlines at $|x| \gtrsim 0.2$ and the horseshoe streamilnes at $|x| \lesssim 0.2$ (Fig.~\ref{fig:hydro}a). The key point in this study is the planetary envelope within the Bondi radius of the planet. The envelope is characterinzed by circular streamlines around the planet, which rotates in the prograde direction due to the Coriolis force. The prograde rotation of the envelope has a significant effect on pebble trajectories (Sect.~\ref{result:trajectory}). Figure~\ref{fig:hydro}b shows the vertical structure of the protoplanet-induced gas flow at $x = 0$. The gas from the disk flows in at high latitudes of the Bondi sphere and exits through the midplane. 
    
    We note that the Bondi radius is smaller than the physical radius of the protoplanet and the protoplanet does not have an envelope, in the cases for the \texttt{m0001-01au} and \texttt{m0003-01au} runs.

    \subsection{Pebble trajectories}
    \label{result:trajectory}
    
    \begin{figure}[tb]
        \centering
        \includegraphics[width=0.9\linewidth, bb = 13 19 557 535]{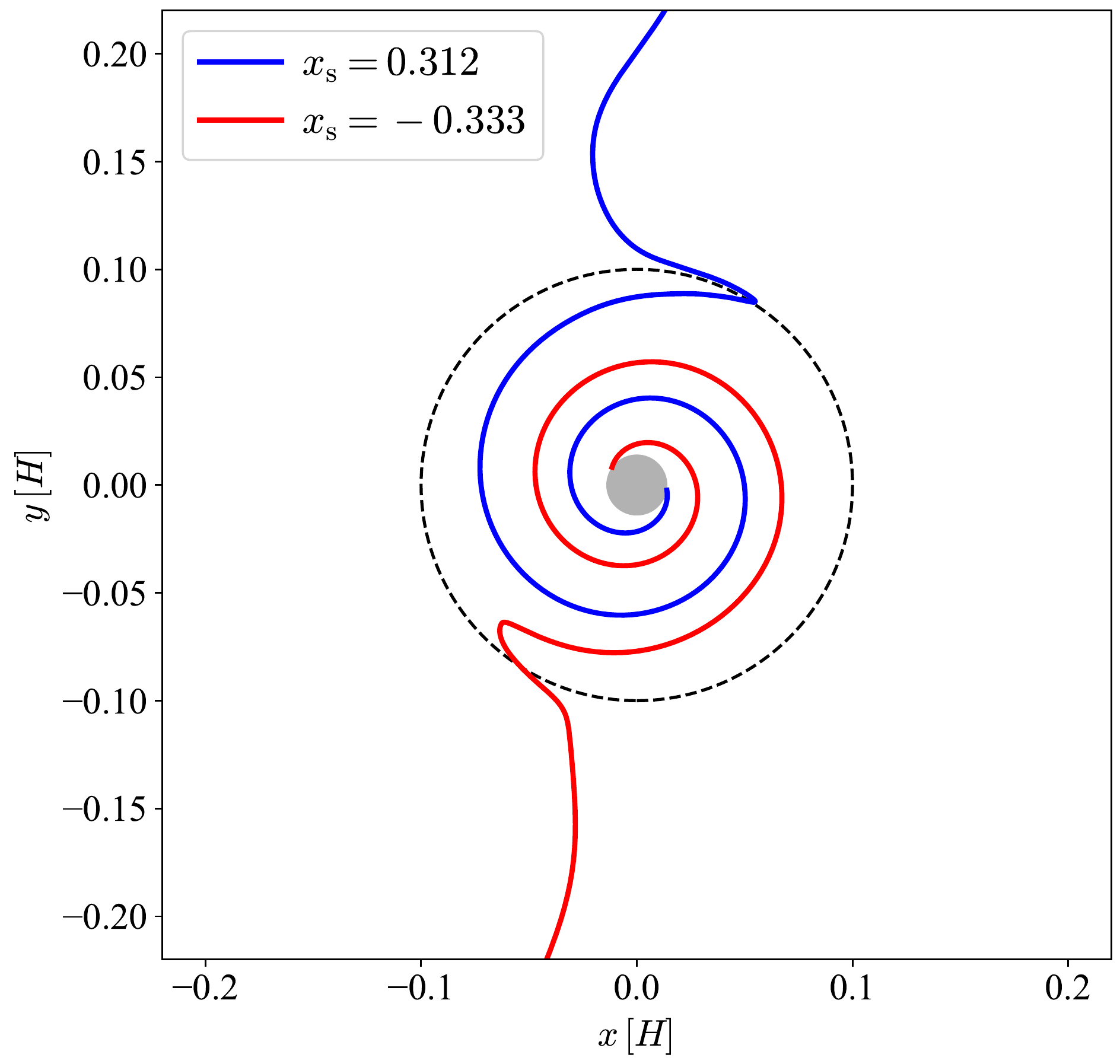}
        \caption{Trajectories of pebbles at the midplane region of the disk influenced by the protoplanet-induced gas flow. We set $z_{\mathrm{s}} = 0$. The blue and red lines correspond to the trajectories of pebbles approaching from the top and bottom of this panel, respectively. The Stokes number of the pebble is set to $\mathrm{St} = 10^{-3}$. The gray circle at the center represents the protoplanet and the black dashed line represents the Bondi radius.}
        \label{fig:orbit}
    \end{figure}
    
    \begin{figure*}[tb]
        \centering
        \includegraphics[width=0.75\linewidth, bb = 11 42 1121 1125]{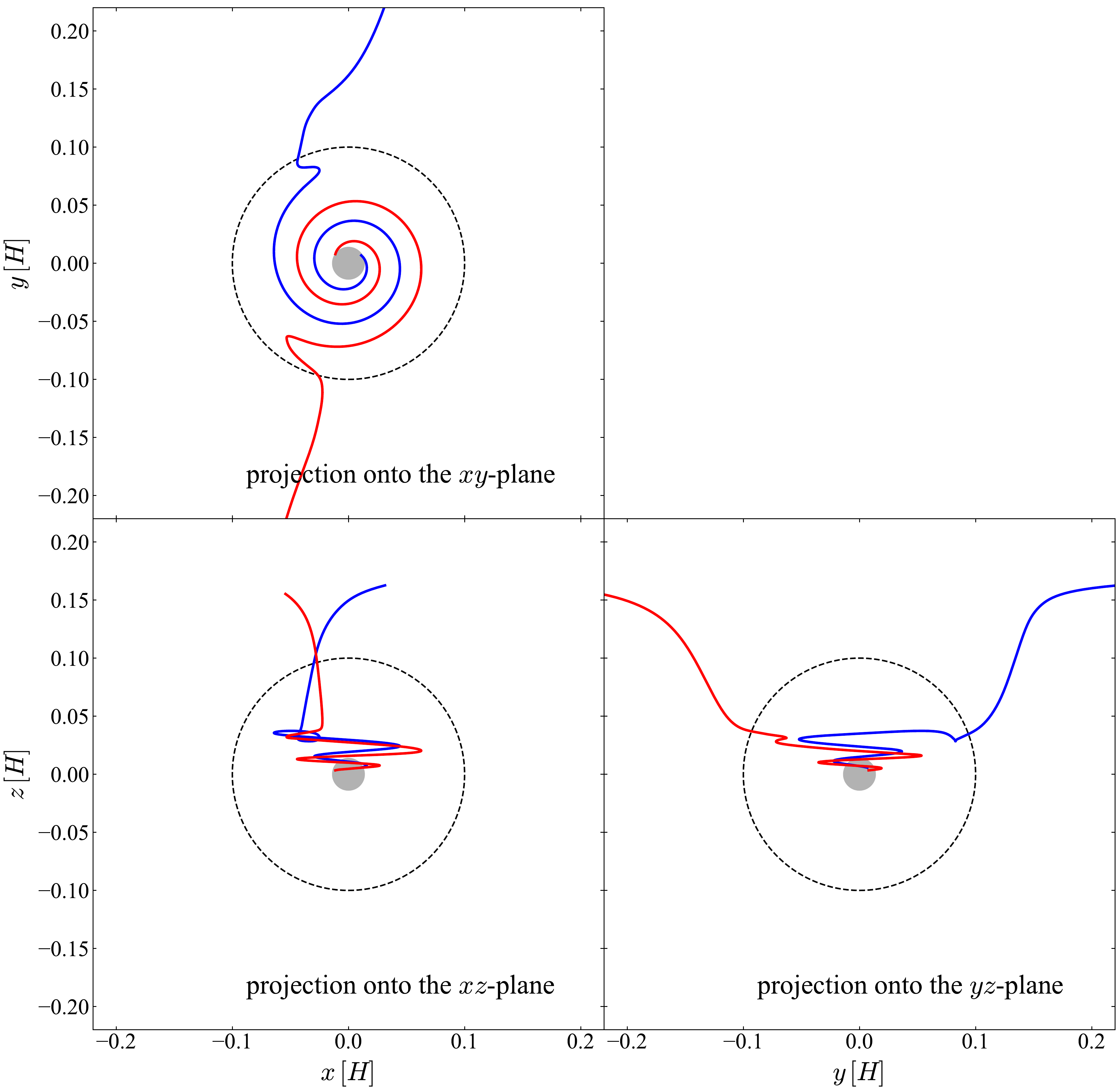}
        \caption{Trajectories of pebbles influenced by the protoplanet-induced gas flow field obtained by the \texttt{m0100-01au} run. These three panels show the same results of orbital calculations, which are projections to the $xy$-, $xz$-, and $yz$-planes, respectively. The blue line represents the trajectory of a pebble with initial position $(x_{\mathrm{s}}, y_{\mathrm{s}}, z_{\mathrm{s}}) = (0.307, 40 \, R_{\mathrm{Hill}}/H, 0.18)$ and the red line represents the trajectory of a pebble with initial position $(x_{\mathrm{s}}, y_{\mathrm{s}}, z_{\mathrm{s}}) = (-0.3275, -40 \, R_{\mathrm{Hill}}/H, 0.18)$. The Stokes number of the pebble is set to $\mathrm{St} = 10^{-3}$. The gray circle at the center represents the protoplanet and the black dashed line represents the Bondi radius.}
        \label{fig:orbit_3D}
    \end{figure*}
    
    We calculate the trajectories of pebbles influenced by the protoplanet-induced gas flow. Figure~\ref{fig:orbit} shows two typical collision orbits at the midplane in the protoplanet-induced gas flow field obtained from \texttt{m0100-01au} run. In Fig.~\ref{fig:orbit}, we assume the Stokes gas drag that is not dependent on the gas density. The Stokes number of pebbles is set to be $\mathrm{St} = 10^{-3}$. To maintain consistency with the hydrodynamical simulation, we assume $\mathcal{M}_{\mathrm{hw}} = 0.03$ for the orbital calculation (Eqs.~(\ref{eq:vx_s}) and (\ref{eq:vy_s})). The blue and red lines represent the trajectories of pebbles coming from $(x_{\mathrm{s}}, y_{\mathrm{s}}, z_{\mathrm{s}}) = (0.312, 40\, R_{\mathrm{Hill}} / H, 0)$ and $(x_{\mathrm{s}}, y_{\mathrm{s}}, z_{\mathrm{s}}) = (-0.333, -40 \, R_{\mathrm{Hill}} / H, 0)$, respectively. Inside the Hill sphere ($R_{\mathrm{Hill}} = 0.32 H$), the pebbles tend to be gravitationally attracted to the protoplanet. The pebbles give prograde spin rotation to the protoplanet,  because they fall onto the protoplanet in counter-clockwise spirals due to the prograde rotation of the envelope.
    
    Figure~\ref{fig:orbit_3D} shows three dimensional views of the trajectories of the pebbles projected to the $x$-$y$, $x$-$z$, and $y$-$z$ planes. The Stokes number of pebbles is $\mathrm{St} = 10^{-3}$. We also assumed the Stokes drag in this orbital calculation. The blue and red lines are trajectories of pebbles coming from $y_{\mathrm{s}} = 40 \, R_{\mathrm{Hill}}/H$ and  $y_{\mathrm{s}} = -40 \, R_{\mathrm{Hill}}/H$, respectively. The initial positions of these pebbles are $(x_{\mathrm{s}}, z_{\mathrm{s}}) = (0.307, 0.18)$ and $(x_{\mathrm{s}}, z_{\mathrm{s}}) = (-0.3275, 0.18)$. The pebbles coming from high altitudes maintain a constant altitude before entering the Hill sphere. This is because the $z$-component of the tidal force is excluded from the simulations (Sect.~\ref{method:EOM}). As shown in the $y$-$z$ plane projection, when the pebbles approach to $|y| \sim R_{\mathrm{Bondi}} / H = 0.1$, they sharply descend toward the protoplanet. After the pebbles are caught in the envelope, they suffer strong drag from the gas envelope and accrete onto the protoplanet (Fig.~\ref{fig:orbit}), which contributes to the prograde spin of the protoplanet.
    
    Orbits that contribute to the prograde rotation are found in broad parameter ranges of $m$ and $\mathrm{St}$. While some orbits transfer retrograde angular momentum to the protoplanet, prograde collisions always dominate, except when the planetary mass is small ($m \ll 1$) and (or) the Stokes number is large ($\mathrm{St} \gtrsim 1$). As $m$ decreases, the size of the envelope decreases. No envelope forms when $R_{\mathrm{Bondi}} \lesssim R_{\mathrm{p}}$, corresponding to $m \la 1.6 \times 10^{-4} (a / 1 \, {\mathrm{au}})^{-3/2}$.\footnote{\cite{Johansen2010} considered the solid body with $m \sim 10^{-6}$ and found that a prograde accretion disk formed around the body under the influence of the strong back-reaction from pebbles to gas. They reported that the prograde accretion disk is compeletely dominated by pebbles, thus the accretion disk in \cite{Johansen2010} is different from the progradely rotating gas envelope considered in this study.} For $\mathrm{St} \ga 1$, pebble motions are not significantly affected by the gas drag.
    
    The envelope-influenced prograde orbits are more pronounced when we adopt the Epstein gas drag, because the gas density increases steeply toward the planetary surface. In the Epstein drag regime, the effective Stokes number is proportional to $\rho_{\mathrm{g}}^{-1}$, leading to the strong gas drag force acting onto the pebbles within the envelope. When $\rho_{\mathrm{g}}$ increases and the mean free path of the gas becomes smaller than the pebble size, the drag law switches from the Epstein to the Stokes regime (Eq.~(\ref{eq:t_stop})). This switch is expected to occur in the vicinity of protoplanets with high gas densities, which limits further increase of the gas drag force. In this study, the switch of the gas drag law is neglected for simplicity. 
    
    We note that pebbles could ablate in the high density, lower envelope \citep{Mordasini2015,Alibert2017}. In this case, pebbles do not directly impact the planetary surface and their angular momentum is deposited to the lower envelope. In this study, assuming that the angular momentum deposited in the lower envelope, which is likely coupled to the solid part of the planet, would be eventually carried to the solid part, we simply integrate the pebble orbits down to the planetary surface without ablation to evaluate the resultant planetary spin (also see Sect.~\ref{discussion:caveat}).

    \subsection{Angular momentum transfer of individual pebbles}
    \label{result:individual_SAM}
    
    \begin{figure*}[tb]
        \centering
        \includegraphics[width=\linewidth, bb = 0 0 720 360]{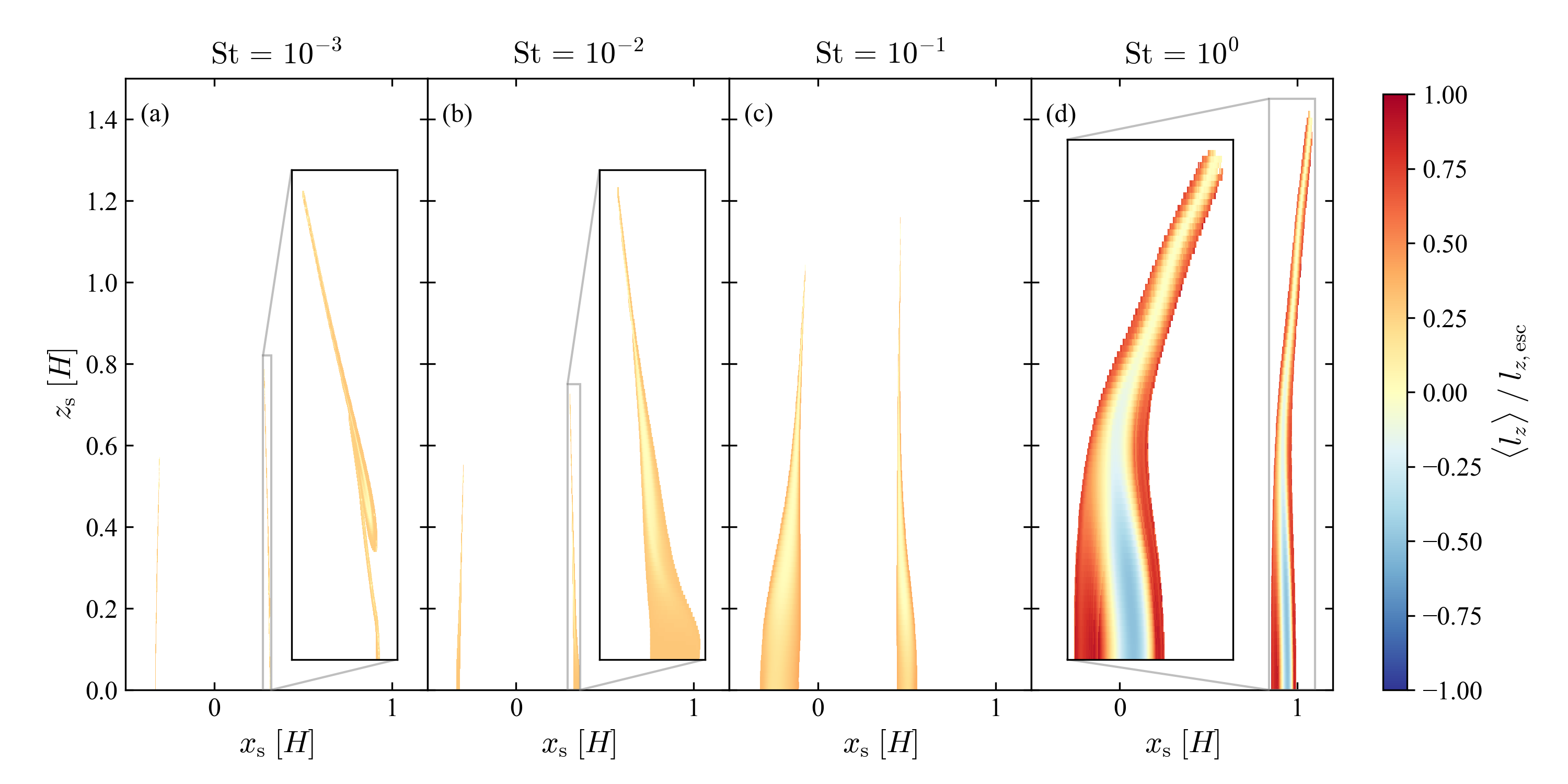}
        \caption{Heat maps of the initial positions of the accreted pebbles with $\mathrm{St} = 10^{-3}$, $10^{-2}$, $10^{-1}$ and $10^{0}$ under the influence of the protoplanet-induced gas flow. The gas flow field is obtained from \texttt{m0100-01au} run. The colors indicate the SAM transferred by individual pebbles to the protoplanet, with red contributing to prograde spin and blue contributing retrograde spin. The two islands in panels a, b, and c correspond to the accretion bands of pebbles coming from the different $y$-directions. Note that due to the low pebble density at the high altitude, the contribution of pebbles coming from high altitude becomes negligible as the Stokes number increases.}
        \label{fig:SAM}
    \end{figure*}

    Figure~\ref{fig:SAM} shows the heat maps of the SAM transferred by accreted pebbles influenced by the protoplanet-induced gas flow obtained from \texttt{m0100-01au} run.  Panels a to d show the results with different Stokes numbers ($\mathrm{St} = 10^{-3} \mathrm{-} 10^{0}$). The red and blue colors indicate the positive and negative SAM that contributes to the prograde and retrograde spins, respectively. When $\mathrm{St} = 10^{0}$, there is no accretion band at $x < 0$ (Fig.~\ref{fig:SAM}d).
    
    As shown in Figs.~\ref{fig:SAM}a and b, where $\mathrm{St} = 10^{-3}$ and $10^{-2}$, pebbles transfer almost constant positive SAM regardless of their initial positions. As mentioned in Sect.~\ref{result:trajectory}, pebbles are strongly affected by the prograde rotation of the envelope and circulate around the protoplanet many times before hitting the planetary surface. During this accretion process, the information on the pebbles' initial positions is lost and the SAM is independent of the initial positions of pebbles.
    
    Figure~\ref{fig:SAM}d, where $\mathrm{St} = 10^{0}$, shows that there are both pebbles that contribute to prograde and retrograde rotations. Pebbles coming from near the edges of the accretion band have positive SAM, while those from the central part of the accretion band have negative SAM. Such orbital patterns are also found by \cite{Visser2020} who investigated the spin of smaller-mass bodies without the envelope ($m \sim 10^{-9} \mathrm{-} 10^{-3}$). When the Stokes number is large, $\mathrm{St} = 10^{0}$, the effect of the prograde envelope on the pebble motion is small, so that the impact points on the planetary surface continuously shift from the prograde side to the retrograde side or vice versa, as the initial positions change.

    \subsection{Net angular momentum transfer}
    \label{result:net_SAM}

    \begin{figure}[tb]
        \centering
        \includegraphics[width=\linewidth, bb = 0 0 720 1152]{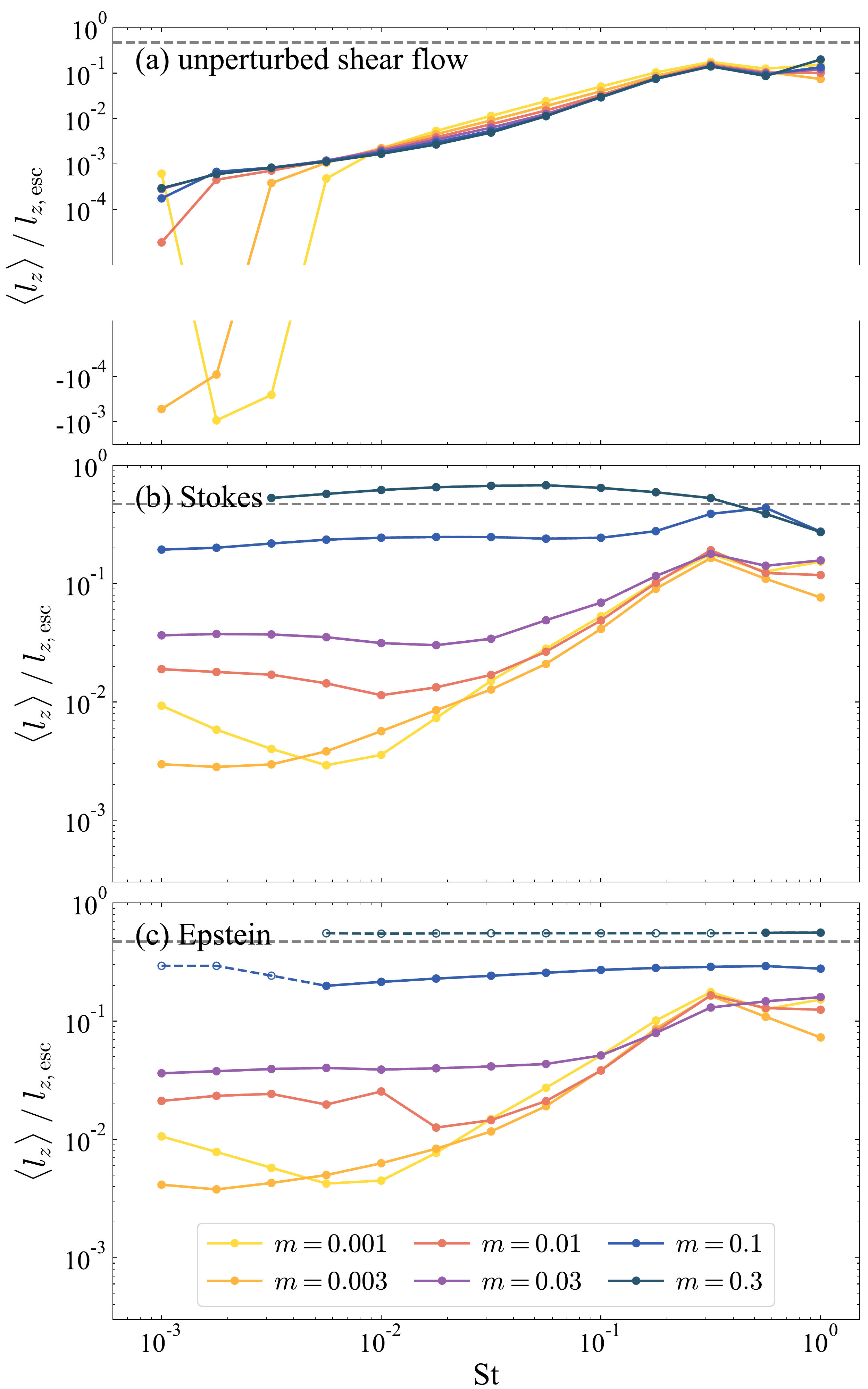}
        \caption{Dependence of the net SAM transferred to the protoplanet on the planetary mass ($m$) and the Stokes number ($\mathrm{St}$), for the unperturbed shear flow with the Stokes drag (\textit{panel a}), the protoplanet-induced gas flow with the Stokes drag (\textit{panel b}), and that with the Epstein drag (\textit{panel c}). The orbital radius is $a = 0.1 \, \mathrm{au}$. Different colors represent different planetary masses ($m$). The solid lines represent cases in which the pebbles accrete over the entire collision band, and the dashed lines represent cases in which some or all pebbles did not accrete and the orbital calculation is interrupted (in other words, dashed lines include the data obtained from Eq.~(\ref{eq:SAM_atm})). The horizontal gray dashed line at the top of each panel represents the net SAM that corresponds to the breakup frequency.}
        \label{fig:dependence_m}
    \end{figure}
    
    Using the data of the SAM transferred by individual pebbles, we calculated the net SAM from Eq.~(\ref{eq:l_z}). Figure~\ref{fig:dependence_m} shows the net SAM transferred to the protoplanet as a function of $\mathrm{St}$ for the different planetary mass, $m$. The orbital radius is $0.1 \, \mathrm{au}$. Figure~\ref{fig:dependence_m}a shows the results with unperturbed shear flow of the gas for a comparison. Figures~\ref{fig:dependence_m}b and c show the results in the planet-induced gas flow where we assumed the Stokes and the Epstein gas drag regimes, respectively. In the Epstein drag case (Fig.~\ref{fig:dependence_m}c), the Stokes number represents the value at the staring position where the gas flow is identical to the unperturbed shear flow. 

    In the case of unperturbed shear flow, the dependence on planetary mass is very weak. The overall trend is that as pebble's Stokes number increases, the net SAM acquired by the protoplanet also increases. When the planetary mass and the Stokes number are small ($m \lesssim 0.003$ and $\mathrm{St} \lesssim 10^{-2}$), the net SAM transferred to the protoplanet has the negative value. The spin rotation is hardly accelerated to exceed the critical breakup frequency represented by the gray dashed line.
    
    We found that the spin of the protoplanets generated by pebble accretion influenced by the protoplanet-induced gas flow is always prograde regardless of the assumed planetary mass and the Stokes number (Figs.~\ref{fig:dependence_m}b and c). We first focus on Fig.~\ref{fig:dependence_m}b where we adopt the Stokes gas drag regime. The striking feature is that for pebbles with $\mathrm{St} \lesssim 0.1$, the net SAM transferred to the protoplanet is an increasing function of the planetary mass. This is because the azimuthal velocity of the gas envelope increases with the planetary mass. The pebbles with small $\mathrm{St}$ spiral onto the planetary surface after they are sufficiently dragged by the gas envelope to acquire prograde rotations (Fig.~\ref{fig:orbit}). The enhancement of the spin rotation is much greater than in the case of unperturbed shear flow. In particular, when $m \gtrsim 0.1$, the expected spin frequency is close to or even higher than the breakup one for a wide range of Stokes numbers. For pebbles with $\mathrm{St} \gtrsim 0.1$, because the effect of gas drag by the envelope is weak, the expected spin is strongly prograde as in the case of the unperturbed shear flow.
    
    In the Epstein gas drag regime, the net SAM transferred to the protoplanet does not differ significantly from that in the Stokes regime (Fig.~\ref{fig:dependence_m}c). Although the effective Stokes number of pebbles decreases as they approach the planetary surface (especially for $m = 0.1$ and $0.3$ with massive envelopes), in reality, the drag law for the pebbles are expected to switch to the Stokes regime in the region where the gas density is sufficiently high. Thus the results shown in Fig.~\ref{fig:dependence_m}c is a limiting case for extremely effective gas drag in the high-density regions. We note that pebbles with the smaller Stokes number circulate around the protoplanet many times in several cases and it takes a very long time for these pebbles to reach the planetary surface. In this study, as described in Sect.~\ref{method:EOM}, we terminate orbital calculations for such pebbles before they reach the planetary surface, in order to reduce the computational cost. For these pebbles, we assume that they would eventually accrete to the protoplanet with the terminal velocity and provide the SAM expressed by
    \begin{equation}
    \label{eq:SAM_atm}
        l_{z, \mathrm{atm}} = r_{\mathrm{in}} \v_{\phi, \mathrm{atm}} + r_{\mathrm{in}}^{2} \Omega_{\mathrm{K}} \, , 
    \end{equation}
    where $\v_{\phi, \mathrm{atm}}$ is the $\phi$ component of the azimuthally averaged gas envelope velocity at the equator of the protoplanet obtained from the hydrodynamical simulation. The parameters for which we used this estimation to calculate the net SAM are shown as dashed lines in Fig.~\ref{fig:dependence_m}.

    \subsection{Dependence on headwind speed}
    \label{result:dependence_hw}
    
    \begin{figure}[tb]
        \centering
        \includegraphics[width=\linewidth, bb = 0 0 720 792]{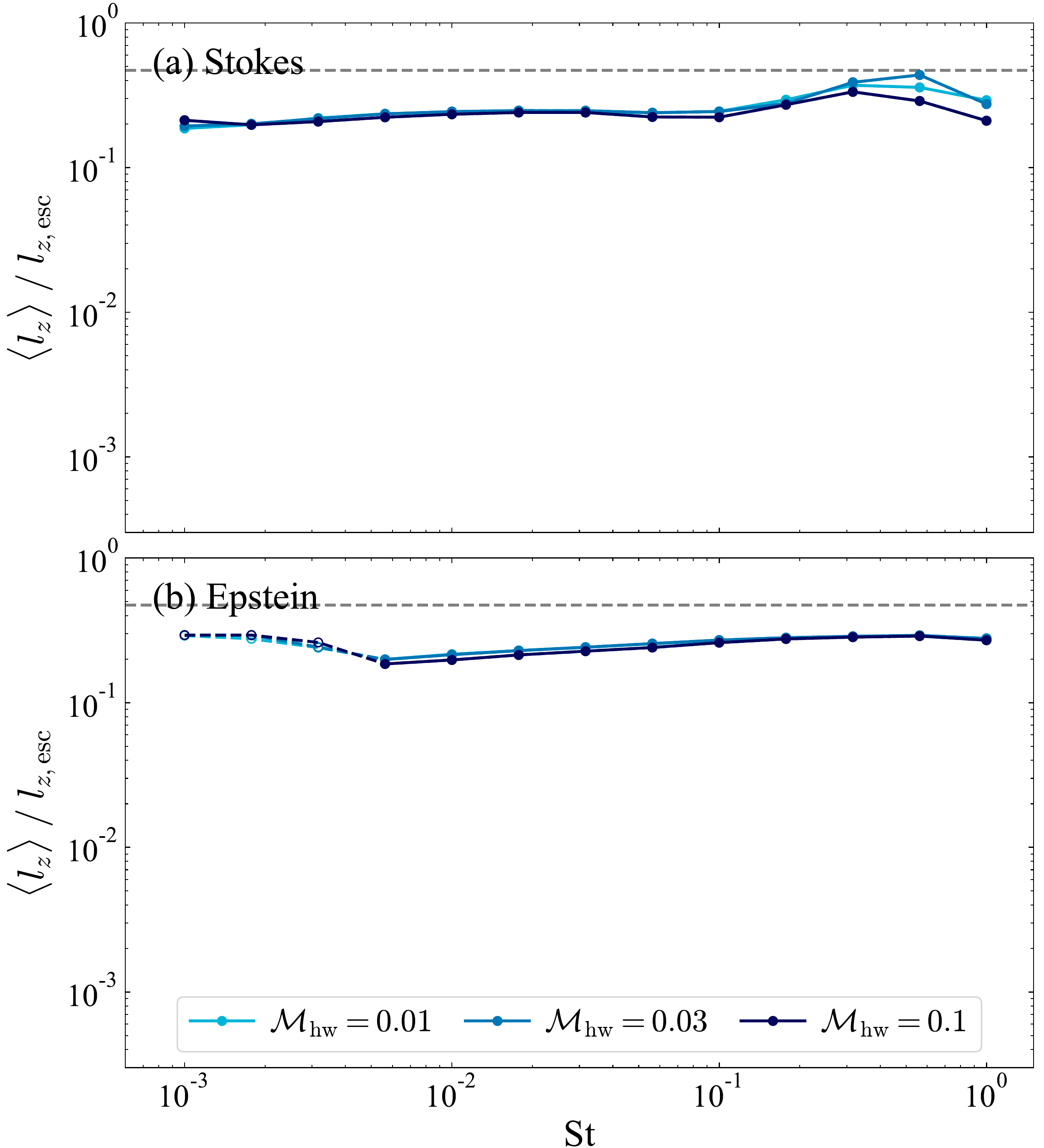}
        \caption{Dependence of the net SAM transferred to the protoplanet on the Mach number of the headwind. The dimensionless mass, orbital radius and smoothing length are $m = 0.1$, $a = 0.1 \, \mathrm{au}$ and $r_{\mathrm{sm}} = 0$, respectively. Different colors represent different Mach numbers. Colored dashed lines represent the data using Eq.~(\ref{eq:SAM_atm}). \textit{Panel a}: the results for the Stokes regime. \textit{Panel b}: the results for the Epstein regime.}
        \label{fig:dependence_hw}
    \end{figure}

    As mentioned in Sect.~\ref{method:EOM}, the basic equation given by Eq.~(\ref{eq:EOM}) includes a simulation parameter of the normalized headwind velocity, $\mathcal{M}_{\mathrm{hw}}$. So far, we have fixed the Mach number of the headwind of the gas, $\mathcal{M}_{\mathrm{hw}} = 0.03$. Figure~\ref{fig:dependence_hw} shows the dependence of the net SAM transferred to the protoplanet on the Mach number of the headwind. These results are obtained from the simulations with $m = 0.1$ at $0.1 \, \mathrm{au}$ in the protoplanet-induced gas flow fields with different $\mathcal{M}_{\mathrm{hw}}$ (\texttt{m0100-01au}, \texttt{m0100-01au-Lhw}, and \texttt{m0100-01au-Hhw}). These results show that the dependence on the headwind is very weak, regardless of the assumed gas drag regime. The results in Fig.~\ref{fig:dependence_hw} for different $\mathcal{M}_{\mathrm{hw}}$ almost completely overlap with each other.
    
    As described in Sects.~\ref{result:trajectory}--\ref{result:net_SAM}, the strong prograde spin is caused by the drag from the prograde rotation of the envelope. Pebbles often circulate around the protoplanets many times, losing the information of the initial conditions. Because the density and velocity of the envelope are hardly affected by the headwind, the $\mathcal{M}_{\mathrm{hw}}$ dependence is very weak.

    \subsection{Dependence on smoothing length}
    \label{result:dependence_rsm}
    
    \begin{figure}[tb]
        \centering
        \includegraphics[width=\linewidth, bb = 0 0 720 792]{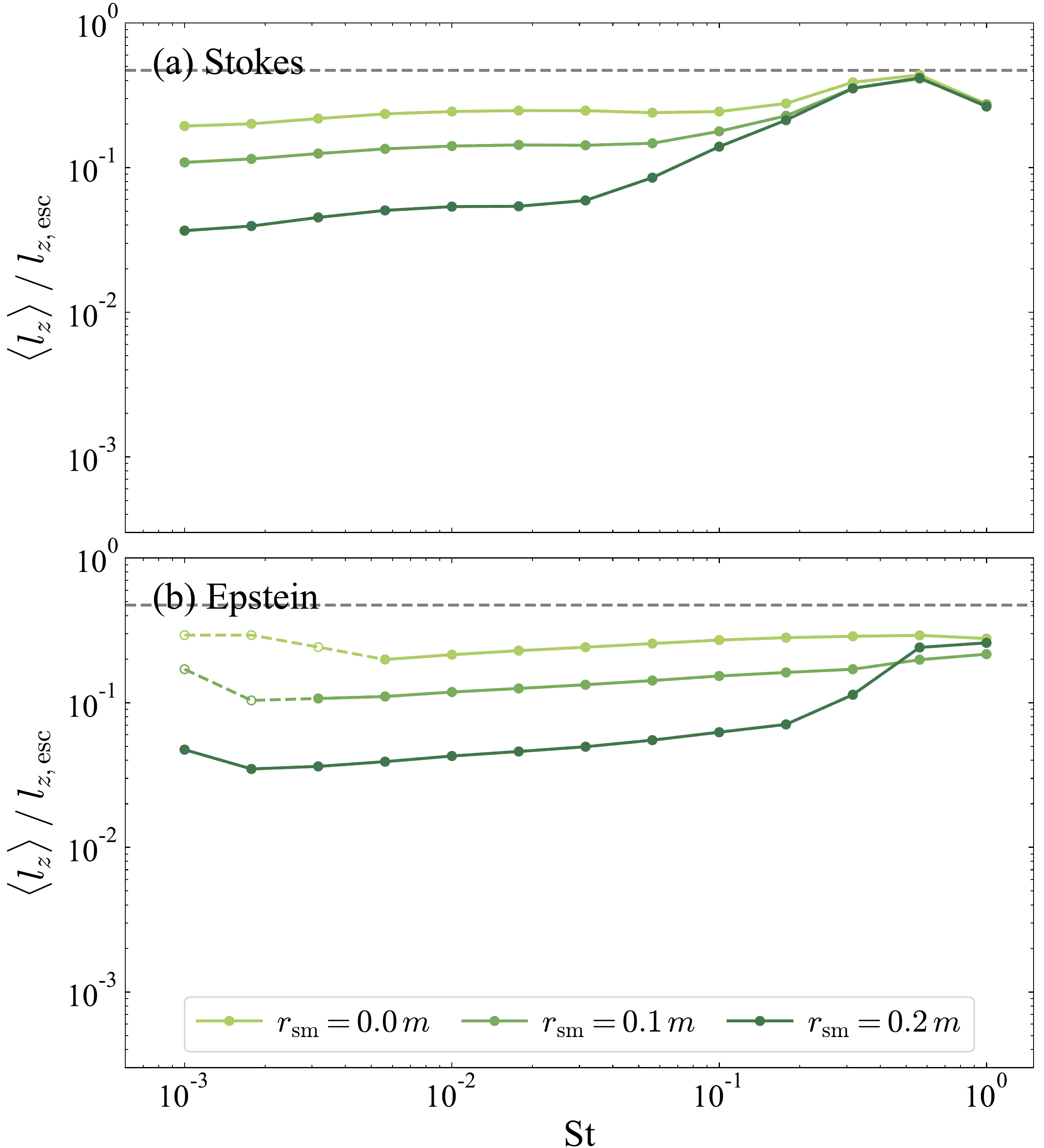}
        \caption{Dependence of the net SAM transferred to the protoplanet on the smoothing length. The dimensionless mass, orbital radius and Mach number of the headwind are $m = 0.1$, $a = 0.1 \, \mathrm{au}$ and $\mathcal{M}_{\mathrm{hw}} = 0.03$, respectively. Different colors represent different smoothing lengths. Colored dashed lines represent the data using Eq.~(\ref{eq:SAM_atm}). \textit{Panel a}: the results for the Stokes regime. \textit{Panel b}: the results for the Epstein regime.}
        \label{fig:dependence_rsm}
    \end{figure}
    
    Ideally, a smoothing length should be set to $r_{\mathrm{sm}} = 0$ to resolve the surface of the planet. So far, we only considered the case of $r_{\mathrm{sm}} = 0$. We confirmed that the azimuthal velocity of the envelope reached the steady state with $r_{\mathrm{sm}} = 0$. However, we found that the small but nonzero inward radial gas flow occurs at the region close to the inner boundary, which means that the hydrostatic equilibrium is not established in this region. This unphysical flow pattern could be eliminated by introducing the smoothing length \citep{Ormel2015b}, but the smoothing for the gravitational potential of the planet would affect the azimuthal velocity of the envelope \citep{Ormel2015b}, and hence affect the SAM transferred to the protoplanet. In this section, we investigate the dependence on the smoothing length.
    
    Figure~\ref{fig:dependence_rsm} shows the dependence on the smoothing length of the planetary gravitational potential (Eq.~(\ref{eq:smoothing})). These results are calculations based on the hydrodynamical simulations with $m = 0.1$ at $0.1 \, \mathrm{au}$ with different smoothing lengths: $r_{\mathrm{sm}} = 0$, $0.1 \, m$, and $0.2 \, m$: \texttt{m0100-01au}, \texttt{m0100-01au-sm01}, and \texttt{m0100-01au-sm02}. The normalized physical radius is $r_{\mathrm{in}} = R_{\mathrm{p}} / H \simeq 0.014$ for $m = 0.1$ at $0.1 \, \mathrm{au}$. We found that as the smoothing length increases, the net SAM decreases. Although the smoothing length is comparable to or smaller than $r_{\mathrm{in}}$, it affects the envelope azimuthal velocity \citep{Ormel2015a}. Larger $r_{\mathrm{sm}}$ reduces the azimuthal velocity of the pebble during its spiral accretion process due to the slower rotation of the envelope gas, resulting in smaller angular momentum transferred to the protoplanet. The weak dependence of the smoothing length for relatively large $\mathrm{St}$ is also consistent with this argument.

    \subsection{Dependence on the orbital radius of the protoplanet}
    \label{result:dependence_au}
    
    \begin{figure}[tb]
        \centering
        \includegraphics[width=\linewidth, bb = 0 0 720 432]{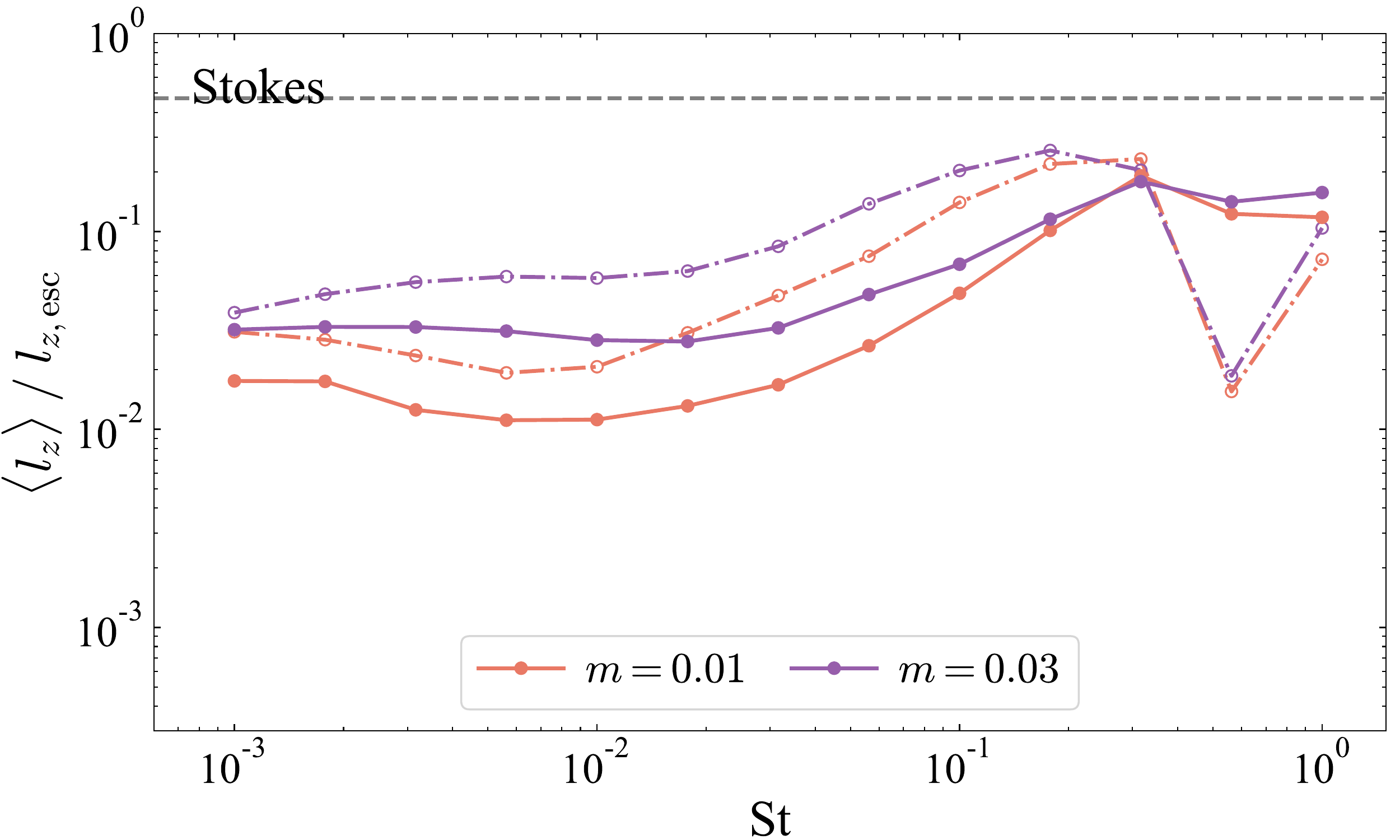}
        \caption{Dependence the net SAM transferred to the protoplanet on the orbital radius of the protoplanet. We adopt the Stokes gas drag regime. The solid and dash-dotted lines correspond to the cases of $a = 0.1 \, \mathrm{au}$ and $1 \, \mathrm{au}$, respectively. The Mach number of the headwind is $\mathcal{M}_{\mathrm{hw}} = 0.03$ and the smoothing length is $r_{\mathrm{sm}} = 0.1 \, m$. Different colors represent different planetary masses.}
        \label{fig:dependence_au}
    \end{figure}
    
    The equation of motion for pebbles is formally parameterized only by $m$ and $\mathrm{St}$ (Eq.~(\ref{eq:EOM})). However, the dimensionless physical radius of the protoplanet, $R_{\mathrm{p}} / H$, depends on $a$ as $R_{\mathrm{p}}/H \propto m^{1/3} / a$ (Eq.~(\ref{eq:Rp1})). Therefore, the $a$-dependence is equivalent to the dependence on the normalized physical radius in this study. As discussed below, $r_{\mathrm{in}} = R_{\mathrm{p}}/H$ affects the rotation velocity of the planetary envelope, and consequently the SAM transferred to the protoplanet.

    Figure~\ref{fig:dependence_au} shows the $a$-dependence of the net SAM transferred to the protoplanet. These results are based on the hydrodynamical simulations with $a= 0.1\,\mathrm{au}$ and $1 \, \mathrm{au}$ for $m = 0.01$ and $0.03$ (\texttt{m0010-01au-sm01}, \texttt{m0010-1au-sm01}, \texttt{m0030-01au-sm01} and \texttt{m0030-1au-sm01} runs). The solid and dash-dotted lines represent the results for $a = 0.1 \, \mathrm{au}$ and $1 \, \mathrm{au}$, respectively. We found that the net SAM generally increases with $a$. This trend, as well as other results shown in the earlier sections, originates from the envelope. When the normalized physical radius of the protoplanet is small, i.e., the orbital distance is large, the gas envelope near the protoplanet's surface rotates in a deeper gravitational potential, leading to the faster rotation of the envelope. Our hydrodynamical simulations show that the azimuthal velocity of the gas envelope is roughly $\v_{\phi,\mathrm{atm}} \propto r^{-1}$. Other literature also shows a similar $r$-dependence \citep{Ormel2015a,Ormel2015b}. Thus, at the planet surface ($r=r_{\mathrm{in}}$), the impact SAM $l_{z} \sim r_{\mathrm{in}} \, \v_{\phi, \mathrm{atm}}$ is independent of $r_{\mathrm{in}}$ for pebbles with $\mathrm{St} \ll 1$. The net SAM transferred to the protoplanet normalized by $l_{z, \mathrm{esc}}$ is $\langle l_{z} \rangle / l_{z, \mathrm{esc}} \propto r_{\mathrm{in}}^{-1/2} \propto a^{1/2}$, because $l_{z, \mathrm{esc}} = \sqrt{2m r_{\mathrm{in}}}$. This explains that the net SAM at $1 \, \mathrm{au}$ is a few times larger than that at $0.1 \, \mathrm{au}$ for $\mathrm{St} \ll 1$ (Fig.~\ref{fig:dependence_au}).\footnote{In the runs with $a = 1 \, \mathrm{au}$ in Fig.~\ref{fig:dependence_au}, the smoothing length is slightly larger than $r_{\mathrm{in}}$ (Table~\ref{table:hydro}). It slightly reduces the difference between the runs at $0.1 \, \mathrm{au}$ and $1 \, \mathrm{au}$ from the theoretically predicted enhancement factor, $\sqrt{10} \sim 3.2$.} Since the influence of the envelope is weaker for the large Stokes number, the differences in ${\bm{\v}}_{\mathrm{g}}$ and $\langle l_{z} \rangle/l_{z, \mathrm{esc}}$ are smaller when $\mathrm{St} \gtrsim 0.1$.

    As described in Sect.~\ref{result:net_SAM}, the spin frequency exceeds the breakup one when $m \gtrsim 0.3$ at $0.1 \, \mathrm{au}$ (Fig.~\ref{fig:dependence_m}). Figure~\ref{fig:dependence_au} and the above argument imply that the spin frequency would reach the breakup one when $m \gtrsim 0.1$ in the case of $1 \, \mathrm{au}$.

\section{Discussion}
\label{discussion}

    \subsection{Pebble isolation mass}
    \label{discussion:isolation}
    
    Our results show that the net SAM transferred from pebbles becomes larger as the protoplanet grows (Fig.~\ref{fig:dependence_m}). Based on the discussion in Sect.~\ref{result:dependence_au}, we introduce the following empirical formula. The spin angular velocity of the protoplanet's rotation exceeds the breakup frequency, when the planetary mass reaches 
    \begin{align}
    \label{eq:m_iso_rot}
        m_{\mathrm{iso,rot}} \sim 0.1 \, \left( \frac{a}{1 \, \mathrm{au}} \right)^{-1/2} \, .
    \end{align}
    Since the centrifugal force due to the fast spin exceeds the planetary gravity near the equator when $\omega > \omega_{\mathrm{crit}}$, the protoplanets would be less likely to grow beyond $m_{\mathrm{iso,rot}}$. We refer $m_{\mathrm{iso,rot}}$ as the rotation-induced isolation mass. From Eq.~(\ref{eq:m_iso_rot}), the dimensional rotation-induced isolation mass can be described by
    \begin{equation}
    \label{eq:M_iso_rot}
        M_{\mathrm{iso,rot}} \simeq 1.2 \,   \,\Biggl( \frac{a}{1 \, \mathrm{au}} \Biggr)^{1/4} M_{\oplus} \, ,
    \end{equation}
    where we assumed the solar mass and luminosity.

    It has been considered that the planetary growth via pebble accretion is inhibited when the planet grows enough to open up a partial gap in a disk. A pressure bump at the outer edge of the gap prevents pebbles from drifting inward to the planet \citep{Lambrechts2014, Bitsch2018, Ataiee2018}. The critical planetary mass for the gap, called the pebble isolation mass ($M_{\mathrm{iso}}$), is estimated to be \citep{Bitsch2018}:
    \begin{equation}
    \label{eq:isolation_mass_1}
        M_{\mathrm{iso,gap}} \simeq 25 \, \left( \frac{H/a}{0.05} \right)^{3} \, \left[ 0.34 \left( \frac{3}{\log_{10} \alpha} \right)^{4} + 0.66 \right] \, M_{\oplus} \, .
    \end{equation}
    Assuming $H/a \simeq 0.033 \, (a / 1 \, \mathrm{au})^{1/4}$ and $\alpha = 10^{-3}$, Eq.~(\ref{eq:isolation_mass_1}) is rewritten as
    \begin{equation}
    \label{eq:isolation_mass_2}
        M_{\mathrm{iso,gap}} \simeq 7.2 \, \left( \frac{a}{1 \, \mathrm{au}} \right)^{3/4} \, M_{\oplus} \, .
    \end{equation}
    In our dimensionless unit, Eq.~(\ref{eq:isolation_mass_2}) can be described by \citep{Kuwahara2020b}: 
    \begin{align}
        m_{\mathrm{iso,gap}} \simeq 0.6 \, .
    \end{align}
    The rotation-induced isolation mass (Eq.~(\ref{eq:m_iso_rot})), $m_{\mathrm{iso,rot}} \sim 0.1 \, (a / 1 \, \mathrm{au})^{-1/2}$, is smaller than the conventional pebble isolation mass for a wide range of the disk ($\gtrsim 0.03 \, \mathrm{au}$). Equation~(\ref{eq:m_iso_rot}) imposes a severe constraint on in situ formation of planets via pebble accretion. For instance, the formation of super-Earths at $a \la 0.1 \, \mathrm{au}$ would be possible only through collisions between plotoplanets after the disk gas dispersal or migration from the outer disk regions after they already grow to the super-Earth sizes.
    
    For $m > m_{\mathrm{iso,rot}}$, the pebbles scattered by the fast spin may stay in the planetary envelope as a planetary ring. The fate of the pebble ring is beyond the scope of this paper and is left for the future work.

    \subsection{Comparison to the planets in the Solar System}
    \label{discussion:observation}
    
    \begin{table*}[tb]
        \centering
        \caption{Orbital radii and masses of terrestrial and icy planets, and the conventional isolation masses and rotation-induced isolation masses at the corresponding orbital radii. The conventional isolation masses and rotation-induced isolation masses are calculated from Eq.~(\ref{eq:isolation_mass_2}) and (\ref{eq:M_iso_rot}), respectively.}
        \scalebox{1}[1]{
        \begin{tabular}{lcccccc}
            \hline
            \hline
            & Mercury & Venus & Earth & Mars & Uranus & Neptune \\
            \hline
            orbital radius, $a$ [$\mathrm{au}$]                                    & 0.39  & 0.72 & 1   & 1.52 & 19.2 & 30.1 \\
            planetary mass, $M_{\mathrm{p}}$ [$M_{\oplus}$]                        & 0.055 & 0.82 & 1   & 0.11 & 14.5 & 17.2 \\
            \hline
            conventional isolation mass, $M_{\mathrm{iso,gap}}$ [$M_{\oplus}$]     & 3.6   & 5.6  & 7.2 & 9.9  & 66   & 93   \\
            rotation-induced isolation mass, $M_{\mathrm{iso,rot}}$ [$M_{\oplus}$] & 0.95  & 1.1  & 1.2 & 1.3  & 2.5  & 2.8  \\
            \hline
        \end{tabular}
        }
        \label{table:planets}
    \end{table*}
    
    We compare our results to the terrestrial and icy planets of the Solar System. To maintain consistency with Sects.~\ref{method} and \ref{result}, we continue to assume the optically-thin limit temperature distribution of the disk around the solar-mass host star (Eq.~(\ref{eq:T_thin})) and discuss the rotation of the planets based on Figs.~\ref{fig:mass} and \ref{fig:dependence_m}.

    We first focus on the terrestrial planets in the Solar System. The rotation-induced isolation mass, $M_{\mathrm{iso,rot}}$, is consistent with the current masses of Earth and Venus, while the current masses of Mercury and Mars are smaller than $M_{\mathrm{iso,rot}}$ by an order of magnitude (Table~\ref{table:planets}). Because the conventional pebble isolation mass ($M_{\mathrm{iso,gap}}$) is too large to be consistent with the Earth and Venus (Table~\ref{table:planets}), an external contingent effect such as the truncation of pebble flux by proto-Jupiter's core must be considered \citep[e.g.,][]{Kruijer2017}. The rotation-induced isolation mass may helpful in explaining the current masses of the Earth and Venus.
    
    Next, we focus on the ice giants. The current masses of Uranus and Neptune are several times larger than $M_{\mathrm{iso,rot}}$ (Table~\ref{table:planets}). Due to the high-speed rotation generated by pebble accretion influenced by the protoplanet-induced gas flow, the growth of the protoplanets would halt at a few Earth masses. These protoplanets may experience giant impacts during disk dispersal, leading to the formation of icy giants in the outer region of the disk. Actually, the spin axis of Uranus and the orbital plane of its moon systems are tilted by 98 degrees, strongly suggesting that Uranus underwent an giant impact in its final formation stage \citep[e.g.,][]{Ida2020}.

    The current spin angular velocity of the Mercury, Venus, Earth, Mars, Uranus and Neptune are $+1.0 \times 10^{-3}$, $-2.5 \times 10^{-4}$, $+5.9 \times 10^{-2}$, $+6.8 \times 10^{-2}$, $-1.7 \times 10^{-1}$ and $+1.6 \times 10^{-1} \omega_{\mathrm{crit}}$ (the positive (negative) value means that the planet has a prograde (retrograde) spin). Because the spin of Mercury may have been influenced by the Solar tide \citep{Colombo1965}, it is not compared with our theoretical prediction. The spin angular momentum of the Earth has been transferred to the Moon's orbital angular momentum by the tidal orbital expansion during $4.5 \, \mathrm{Gyr}$, the total angular momentum of the Earth-Moon system should be considered. The converted Earth-Moon's effective angular velocity is $\simeq 0.29 \, \omega_{\mathrm{crit}}$. 

    Our result predicts $\omega \sim \omega_{\mathrm{crit}}$ for the Earth. It is high enough to achieve the large angular momentum of the Earth-Moon's angular momentum, although $2/3$ of the angular momentum must be lost by some mechanisms to be compabale to the current value. For Mars, $\omega \sim 0.1 \omega_{\mathrm{crit}}$ is predicted, which is consistent with the current Mars spin.
    
    The major problem in our results is that a planet's rotation cannot be retrograde. Several bodies in the Solar System rotate in the retrograde direction, such as Pallas, Hygiea, and Venus \citep{Warner2009}. The existence of retrograde-rotating asteroids ($\lesssim 500 \, \mathrm{km}$) can be explained by the absence of the envelope \citep{Visser2020}. For the case of Venus, the influence of the envelope should be considered, but Venus could be marginally influenced by the Solar tide \citep{Laconte2015}, which is not considered in this study. Although the nonisothermal hydrodynamical simulations with the viscosity of the fluid and the dust opacity show a weak retrograde motion of the gas within the Hill sphere \citep{Lambrechts2017}, considering the effect of the viscosity or opacity is beyond the scope of this study.

    \subsection{Moon formation}
    \label{discussion:moon}
    
    Our results indicate that an Earth-mass planet at $\sim 1 \, \mathrm{au}$ has fast spin rotation if it is formed through pebble accretion influenced by the protoplanet-induced gas flow, which has an interesting implication for the formation of the Moon. The giant impact hypothesis is the current standard model for the Moon formation \cite[e.g.,][]{Benz1986, Ida1997, Canup2001}. Because the giant-impact hypothesis assumes that an oblique impact by a Mars-sized body, it also explains the large angular momentum of the Earth-Moon system. However, it has a difficulty to account for the Moon's stable isotope ratios that are identical to the Earth's ones even with the up-to-date high resolution measurements \cite[e.g.,][]{Wiechert2001}. The SPH simulations of giant impacts predict that the Moon is composed mainly of materials from the impactor \citep{Canup2001, Canup2004, Canup2008}, which is expected to have different isotope ratios from those of the Earth.
    
    To reconcile this inconsistency, \cite{Cuk2012} proposed a model that combines the giant-impact hypothesis and the classical fission hypothesis. If a Ganymede to Mercury mass body impacts the primordial Earth that was already spinning fast with a period as short as 2.3 hours (corresponding to $\omega \simeq 0.61 \, \omega_{\mathrm{crit}}$), the Moon is formed mainly from the Earth's mantle in a fission-like manner. Our results suggest that the proto-Earth could have acquired such a sufficiently fast rotation by pebble accretion, providing the initial state of \cite{Cuk2012}'s model.

    \subsection{Caveats}
    \label{discussion:caveat}
    
    In our simulations, we ignore several physical processes for simplicity. Here we list up a few physical processes that could affect our results.
    
    The first point is the back-reaction from the pebbles to the gas. Because the gas flow was obtained by the independent hydrodynamical simulations, the back-reaction was neglected. We have shown that pebble motions are significantly changed in the planetary envelope when the Stokes number of the pebbles is small. It implies that a large amount of angular momentum is transferred from the gas envelope to the pebbles and the envelope rotation may be slowed down. The effect of the back-reaction depends on the efficiency of the atmospheric recycling. In our nonisothermal hydrodynamical simulations, the envelope is isolated from the surrounding disk gas, which is consistent with the previous studies \citep{Cimerman2017, Lambrechts2017, Kurokawa2018}. The isolation of the envelope is caused by the buoyancy, which is originated from the entropy gradient \citep{Kurokawa2018}. This suggests that the atmospheric recycling is inefficient, but the efficiency of the atmospheric recycling under nonisothermal condition is a controversial issue \citep{Moldenhauer2021, Moldenhauer2022}. Under isothermal condition, efficient recycling is allowed due to the absence of buoyancy \citep{Ormel2015b, Fung2015, Kuwahara2019, Bethune2019}. If the recycling is fast enough, the effect of the back-reactions can be neglected.

    Second, we did not consider viscosity of the gas and associated angular momentum exchange between the gas and the protoplanet. A protoplanet grown by pebble accretion would be expected to already have relatively fast rotation \citep{Visser2020, Visser2022}. The gas velocity field around the protoplanet may change depending on the extent to which the envelope rotation is locked to the protoplanet's spin. This effect may only occur fairly close to the planet's surface, but it could be an important factor because it is the last part of pebbles' orbits before accretion.
    
    The third is the ablation that we have already mentioned in Sect.~\ref{result:trajectory}. While the angular momentum exchange between the pebbles and the envelope through gas drag is large in the most upper-layer near the Bondi radius as shown in Fig.~\ref{fig:orbit}, the deposited angular momentum would be transferred to disk regions outside the Bondi radius. On the other hand, the angular momentum deposited to the bottom layer, where pebbles would suffer ablation, is likely to be transferred to the solid part of the planet rather than to the disk region outside the Bondi radius. It suggests that the effect of ablation is negligible. However, more detailed studies are needed for this point.

\section{Summary}
\label{conclusion}

    We have investigated how pebble accretion induces the spin of a protoplanet under the influence of the protoplanet-induced gas flow. We performed 3D hydrodynamical simulations of the gas flow around the protoplanet in a local frame. Using the simulated gas velocity (and gas density) field, we numerically integrated the equation of motion of pebbles. We calculated the spin angular momentum per unit mass transferred from individual pebbles to the protoplanet at the collisions. The main results are summarised as follows.
    
    \begin{enumerate}
    
        \item An isolated envelope forms around a protoplanet with $\gtrsim 10^{3} \, \mathrm{km}$ in size or $\gtrsim 10^{-3} M_\oplus$ in mass, which rotates in the prograde direction due to the Coriolis force. Pebbles are dragged by the envelope and usually end up with prograde oblique impacts to the protoplanet. 
        
        \vskip.8 \baselineskip
        
        \item The protoplanet acquires the prograde rotation via pebble accretion influenced by the protoplanet-induced gas flow, regardless of the assumed planetary mass, the Stokes number, the Mach number of the headwind, and the orbital distance of the protoplanet. This is because the prograde contribution from pebbles dominates due to the prograde rotation of the envelope.   
        
        \vskip.8 \baselineskip
        
        \item As planetary mass increases, the density and rotation velocity of the envelope increase, resulting in greater SAM transfer to the protoplanet. The spin frequency of the protoplanet would exceed the breakup frequency when the dimensionless thermal mass of the planet reaches the rotation-induced isolation mass, $m_{\mathrm{iso,rot}} \simeq 0.1 \, (a / 1 \, \mathrm{au})^{-1/2}$, suggesting that the protoplanet does not grow any further via pebble accretion. The rotation-induced isolation mass could be significantly smaller than the conventional pebble isolation mass.
        
        \vskip.8 \baselineskip
        
        \item It is a robust conclusion that an Earth-mass planet at $a \sim 1 \, \mathrm{au}$ acquires fast prograde spin with near the breakup angular velocity ($\omega_{\mathrm{crit}}$) if the planet grows predominantly through pebble accretion under the influence of the protoplanet-induced gas flow. The Earth's mass is consistent with $m_{\mathrm{iso,rot}}$ and the predicted spin frequency is comparable to $\omega_{\mathrm{crit}}$, allowing the Moon formation in a fission-like manner.
        
    \end{enumerate}

\begin{acknowledgement}
    We are grateful to an anonymous referee for a very careful and constructive review. We thank Athena++ developers: James M. Stone, Kengo Tomida and Christopher White. This work has profited immensely from discussion with Takayuki Tanigawa, Takanori Sasaki and Satoshi Okuzumi. Numerical computations were in part carried out on Cray XC50 at the Center for Computational Astrophysics at the National Astronomical Observatory of Japan. This work was supported by JSPS KAKENHI Grant numbers 20J20681 and 21H04512.
\end{acknowledgement}

\bibliography{bibfilename}

\end{document}